\documentclass[11pt]{article}
\usepackage{mathtools}
\usepackage{fullpage}
\usepackage{graphicx}
\usepackage{multirow}
\usepackage{amsmath}
\usepackage{amssymb}
\usepackage[linesnumbered, ruled]{algorithm2e}
\usepackage{hyperref}
\usepackage[doi=true,isbn=false,url=false,sorting=none, style=numeric]{biblatex}
\renewbibmacro{in:}{}
 \usepackage{setspace}
\usepackage{amsthm} 
\usepackage[dvipsnames]{xcolor} 
\usepackage{cases}
\usepackage[dvipsnames]{xcolor}
\newcommand{\xnew}{{X_{\text{new}}}}
\newcommand{\xnewt}{\xnew^\top}

\DeclareMathOperator{\sd}{sd} 
\DeclareMathOperator{\ci}{CI} 
\DeclareMathOperator{\ind}{\mathbb{I}}

\newcommand{\R}{\mathbb{R}}

\newcommand{\bvec}[1]{\boldsymbol{#1}}
\newcommand{\N}{\mathcal{N}}
\DeclareMathOperator{\var}{Var}

\title{An Adaptively Resized Parametric Bootstrap\\for Inference in High-dimensional Generalized Linear Models\\}

  \author{Qian Zhao\thanks{Department of Biomedical Data Science, Stanford
    University, Stanford, CA 94305} \and 
   \and Emmanuel J. Cand\`es\thanks{Department of Mathematics, Stanford
    University, Stanford, CA 94305} \thanks{Department of Statistics, Stanford
    University, Stanford, CA 94305} }
    
\begin{document}

\maketitle

\begin{abstract}				
  Accurate statistical inference in logistic regression models remains
  a critical challenge when the ratio between the number of parameters
  and sample size is not negligible. This is because
  approximations based on either classical asymptotic theory or
  bootstrap calculations are grossly off the mark.  This paper
  introduces a resized bootstrap method to infer model parameters in
  arbitrary dimensions. As in the parametric bootstrap, we resample
  observations from a distribution, which depends on an
  estimated regression coefficient sequence. The novelty is that this
  estimate is actually far from the maximum likelihood estimate
  (MLE). This estimate is informed by recent theory studying
  properties of the MLE in high dimensions, and is obtained by
  appropriately shrinking the MLE towards the origin.  We demonstrate
  that the resized bootstrap method yields valid confidence intervals
  in both simulated and real data examples. Our methods extend to
  other high-dimensional generalized linear models.
\end{abstract}

\section{Introduction}

The bootstrap is a well-known resampling procedure introduced in
Efron's seminal paper \cite{efron1979bootstrap} for approximating the
distribution of a statistic of interest. Its popularity stems from a
combination of several elements: it is conceptually rather
straightforward; it is flexible and can be deployed in a whole suite
of delicate inference problems \cite{efron1981censored,efron1985parametric,efron1996phylogenetic}; and finally, whenever theoretical calculations are
impossible, the bootstrap often provides an excellent approximation to
the distribution under study. As a result, researchers from a
spectacular array of disciplines have used the bootstrap for
hypothesis testing \cite[Chapter~1.8]{subsampling_book}, model
selection \cite{shao1996bootstrap_model_selection}, density estimation
\cite{franke1992spectral}, and many other important statistical
inference problems.

The bootstrap can usually be understood via the plug-in principle
\cite[Chapter~4]{intro_bootstrap_efron_tibshirani}. Suppose we observe
${X}_i\in\R^p$, $i=1,\ldots, n$, sampled independently and identically from a distribution $F$.
We wish to infer the distribution of a statistic
$t_F({X}_1, {X}_2,\ldots, {X}_n)$, which can be a complicated
functional of the data aimed at estimating the number of modes $F$
has. For instance, we may be interested in the 90\% quantile of
$t_F({X}_1,\ldots, {X}_n)$.\footnote{The subscript $F$ means that the
  ${X}_i$'s are i.~i.~d.\ samples from $F$.} The plug-in principle
estimates the distribution of $t_F({X}_1,\ldots, {X}_n)$ by that of
$t_{\hat{F}}({X}_1^*,\ldots, {X}_n^*)$, wherein $\hat{F}$ is an
estimate of $F$, and $({X}_1^*,\ldots, {X}_n^*)$ is a draw from
$\hat F$. In other words, by resampling observations from $\hat F$, we
obtain a distribution we hope closely resembles that of
$t_F(X_1, \ldots, X_p)$.

\newcommand{\goto}{\rightarrow}

Naturally, statisticians have since the beginning studied the accuracy
of the bootstrap. Broadly speaking, the bootstrap is known to be
consistent, i.e.,
$t_{\hat{F}}(X^*_1,\ldots, X^*_n)\stackrel{}{\longrightarrow}
t_F(X_1,\ldots, X_n)$ in distribution, under the conditions that (1) the distribution of
$t_F({X}_1, {X}_2,\ldots, {X}_n)$ varies smoothly near $F$, and (2) 
$\hat{F}$ converges to $F$ (See \cite{bickel1981},
\cite[Section~3.1]{diciccio1988} and
\cite[Section~1.2]{subsampling_book}). The second condition is
typically satisfied for appropriately chosen estimates $\hat{F}$
whenever the data dimension $p$ is fixed. In addition to general
theory, statisticians have carried out detailed studies for specific
statistics including the sample mean \cite{bickel1981, hall1992},
regression coefficients \cite{shorak1982bootstraprobustregression,
  bickelfreedman1982,bickel1981, mammen1993bootstrap}, and continuous
functions of the empirical measure \cite{gine1990bootstrap}, and so
forth.

Motivated by the abundance of high-dimensional data, researchers are
increasingly studying statistical methods in the {\em high-dimensional
  setting} in which the number of variables $p$ grows with the number
of observations $n$. Specifically, this article concerns the accuracy
of bootstrap methods when $p$ and $n$ are both very large and perhaps
grow with a fixed ratio. In linear regression for example, while
  the residual bootstrap is weakly consistent if $p$ is fixed and
  $n \goto \infty$, it is inconsistent when $n, p \goto \infty$ in
  such a way that $p/n \goto \kappa > 0$; to be sure,
  \cite{bickelfreedman1982} displays a data-dependent contrast, i.e., a linear
  combination of coefficients, for which the estimated contrast
  distribution is asymptotically incorrect. Motivated by results from
high-dimensional maximum likelihood theory
\cite{elkaroui2013,elkaroui2013p2,elkaroui2018},
\cite{elkaroui2018boot} proposed to use corrected residuals to achieve
correct inference. Another example is this: although the
  nonparametric bootstrap can be used to construct a valid confidence
  region for the spectrum of a covariance matrix when the problem
  dimension is fixed \cite{beran1985covariance,eaton1991bootstrap}, it
  yields incorrect estimates of the distribution of the largest
  eigenvalue if $p/n \goto \kappa > 0$
  \cite{karoui2016bootstrapcovariance}. With
  the exception of these two studies, the accuracy of the bootstrap in
  other high-dimensional problems has not been much researched.

In this paper, we study the bootstrap for inferring the distribution
of the maximum likelihood estimator (MLE) in high-dimensional logistic
regression models. We find that the standard parametric bootstrap and
the pairs bootstrap are both incorrect (Section
\ref{sec:intro_example}), a finding which echoes with
\cite{elkaroui2018boot}. We also show that recent high-dimensional
maximum likelihood theory (HDT) developed for multivariate Gaussian
covariates does not correctly predict the distribution of the MLE when
the covariates are heavy tailed; this is analogous to findings in
\cite{elkaroui2018}. Both these failures call for solutions and in
this paper, we design a novel resized bootstrap by combining the
bootstrap method with insights from HDT. We demonstrate that the
resized bootstrap yields confidence intervals attaining nominal
coverage regardless of the covariate distribution. Finally, we extend
our methods to other generalized linear models.


\subsection{High-dimensional maximum likelihood
  theory} \label{sec:high_dim_theory}

We begin by briefly reviewing recent theory about M-estimators in the
high-dimensional setting in which both the number of observations $n$
and the number of variables $p$ go to $\infty$ while the ratio $p/n$
approaches a constant $\kappa > 0$. This theory---from now on, we use
HDT as a shorthand for high-dimensional theory---generalizes the
classical asymptotic setting, and offers a more accurate
characterization of the distribution of M-estimators when both $n$ and
$p$ are large.  In particular, a considerable amount of research has
studied the behavior of M-estimators in high-dimensional regression
and penalized regression \cite{elkaroui2013, elkaroui2013p2,
  donoho2016, vandergeer2014lassoCI, zhang2013lassoCI,
  bellec2021asymptotic, celentano2020lasso}.  

Consider a logistic model in which the covariates $X \in \R^p$ are
mutivariate Gaussian and $\mathbb{P}(Y = 1|X) = \sigma(X^\top \beta)$, where
$\sigma(t) = 1/(1 + e^{-t})$ is the usual sigmoid function. Previous research \cite{zhao2020} showed that if $\hat \beta$ denotes the MLE, then   
\begin{equation}\label{eq:gaussian_asym_theory}
        \frac{\sqrt{n}(\hat{\beta}_j - \alpha_\star \beta_j)}{\sigma_\star / \tau_j} \stackrel{d}{\longrightarrow} \N(0,1),
\end{equation}
where $\beta_j$ (resp.~$\hat \beta_j$) is the $j$th (resp.~estimated)
model coefficient. In contrast to classical asymptotic theory, which
states that the MLE is unbiased, the MLE is centered at
$\alpha_\star \beta_j$, for some $\alpha_\star > 1$ whenever $\kappa$
is positive. The standard deviation is $\sigma_\star/ \tau_j$; here,
$\tau_j$ is the conditional standard deviation of the $j$th variable
given all the other variables whereas the parameters $\alpha_\star$
and $\sigma_\star$ are determined by $\kappa$ and the signal strength
$\gamma^2$ defined as $\gamma^2 = \var({X}^\top
{\beta})$. The parameters $\alpha_\star$ and $\sigma_\star$ both increase as either
the dimensionality $\kappa$ increases or the signal-to-noise ratio
$\gamma$ increases \cite[Figure~7]{sur2019p2}. (To be complete, we
stress that Eqn.\ \eqref{eq:gaussian_asym_theory} holds with the
proviso that the magnitude of $\beta_j$ is not extremely large; we
refer to the reader to \cite{zhao2020} for a quantitative description.) 
 
The approximation \eqref{eq:gaussian_asym_theory} happens to be very
accurate for moderately large sample sizes \cite{zhao2020}, e.g., when
$n = 4000$ and $p=400$, and is accurate for relatively small sample
sizes, i.e., $n=200$ and $p=20$ \cite[Appendix~G]{sur2019p2}. Further, 
\eqref{eq:gaussian_asym_theory} is expected to hold for 
sub-Gaussian covariates, see  \cite{zhao2020} for 
empirical studies supporting this claim.

Having said all of this, \eqref{eq:gaussian_asym_theory} does not hold
when the covariates follow a general distribution. For instance,
suppose the covariates $X\in\R^p$ are sampled from a multivariate
$t$-distribution. Then we expect that $\alpha_\star$ and
$\sigma_\star$ would depend on the degrees of freedom of the
$t$-distribution.  In this direction, \cite{elkaroui2018} studied ridge
regression in linear models when the covariates follow a multivariate
$t$-distribution, and proved that the variance of the ridge estimate
does depend on the geometry of the covariates. Asides from this, we
know very little about the distribution of M-estimators when
covariates come from an arbitrary distribution.

\subsection{An example with non-Gaussian covariates}\label{sec:intro_example}
Having succinctly described the high-dimensional theory, we simulate a
high-dimensional logistic regression model with 4000 observations and
400 covariates ($n = 4000$ and $p = 400$). We sample covariates from a
multivariate $t$-distribution and standardize each variable so that
$\var(X_j) = 1/p$. We pick 50 non-null variables and sample their
coefficients from a mixture of Gaussians $\N(5, 1)$ and $\N(-5, 1)$
with equal weights.

\begin{figure}[ht]
    \centering
    \includegraphics[width = 0.6\textwidth]{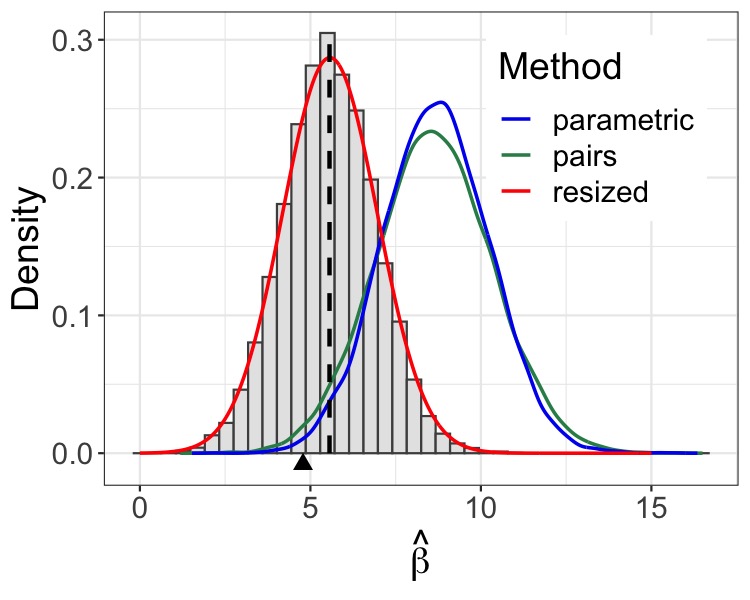}
    \caption{Histogram of the logistic MLE of a randomly chosen
      coefficient in 10,000 repeated experiments. Here, the covariates
      are sampled from a multivariate $t$-distribution with 8 degrees
      of freedom. The bootstrap MLE densities are displayed for the
      parametric bootstrap (blue), the pairs bootstrap (green) and the
      proposed resized bootstrap (red). The triangle indicates the
      true coefficient and the dashed line indicates the average
      MLE. }
    \label{fig:intro_ex}
\end{figure}

Figure \ref{fig:intro_ex} presents a histogram of a coordinate of the
MLE from repeated experiments. From the bell-shaped curve, we conclude
that the MLE is approximately Gaussian. Although the value of the true
coefficient under study is 4.78, the average MLE is 5.56, which shows
that the MLE is biased upward and the inflation factor is roughly
equal to $\alpha_j = \bar{\beta}_j / \beta_j = 1.16$. The empirical
standard deviation (std.~dev.) of the MLE is equal to 1.34; however,
the classical theory estimates that the std.~dev.\ equals 1.15. We thus
see that because of both a poor centering and a poor assessment of
variability, the classical Wald confidence interval would
significantly undercover $\beta_j$. Now HDT from Section
\ref{sec:high_dim_theory} estimates the bias to be
$\alpha_\star = 1.14$ and the standard deviation to be
$\sigma_\star/\tau_j=1.25$. This implies that while capturing the
bias, HDT slightly underestimates the std.~dev.\ of the MLE.

Next, we apply the parametric bootstrap and pairs bootstrap and
display in Figure \ref{fig:intro_ex} the density curves of the
bootstrap MLEs from one experiment.
\begin{itemize}
  \item For the parametric bootstrap, we
generate samples by fixing the covariates at the observed values and
sample responses from a logistic model whose coefficients equal the
MLE; put another way, we choose $\hat{F}={F}_{\hat{\beta}}$. The
parametric bootstrap (blue) does not begin to describe the MLE
distribution since the average value is 8.68, about twice that of the
true coefficient, and the std.~dev.\ is 1.55.
\item The pairs bootstrap
generates bootstrap samples by sampling with replacement from the
observed data, i.e., we choose $\hat{F}$ to be the empirical
distribution. The pairs bootstrap also fails to approximate the MLE
distribution since the green curve shifts to the right and is much
wider than the histogram (mean is 8.63 and std.~dev.\ 1.71).
\end{itemize}

Finally, the red curve in Figure \ref{fig:intro_ex} shows the accuracy
of the proposed resized bootstrap.  We can see that this best
describes the MLE distribution; for instance, both the mean (5.54) and
standard deviation (1.39) are close to the true values.

\section{Why does the bootstrap fail?}\label{sec:why_bootstrap_fails}

The pairs bootstrap fails in the high-dimensional setting because it
effectively inflates the dimensionality ratio $\kappa = p/n$. In
particular, when $n$ is large, the number of unique pairs
$(X_i^*, Y_i^*)$ in a bootstrap sample is approximately $(1-1/e) n$ on
average \cite{mendelson2016}. Consequently, the effective
dimensionality ratio $\kappa e/(e-1)$ in the bootstrap sample is 
larger than $\kappa$. Because the bias and variance of the MLE
increase as $\kappa$ increases \cite[Figure~7]{sur2019p2}, the pairs
bootstrap tends to over-estimate both the bias and standard error.

While the pairs bootstrap over-estimates $\kappa$, the parametric
bootstrap fails because the signal strength $\gamma$ is inflated in
the bootstrap samples. Suppose for simplicity that the covariates are
independent $\N(0,1)$. Then \cite[Theorem~2]{sur2019p2} shows that
\begin{equation}\label{eq:signal_strength_mle}
     \lim_{n,p\to\infty}\var({X}_{\text{new}}^\top \hat{{\beta}}) \stackrel{a.s.}{=} \alpha_\star^2 \gamma^2 + \kappa \sigma_\star^2 > \gamma^2,
\end{equation}
whereas $\var(X^\top \beta) = \gamma^2$. Here, $X_{\text{new}}$ is a
new random sample independent from the training set. Because a higher
$\gamma$ leads to higher bias and variance \cite[Figure~7]{sur2019p2},
the parametric bootstrap also tends to over-estimate the bias and
standard error of the MLE.

\begin{figure}[ht]
    \centering
    \includegraphics[width = 0.5\textwidth]{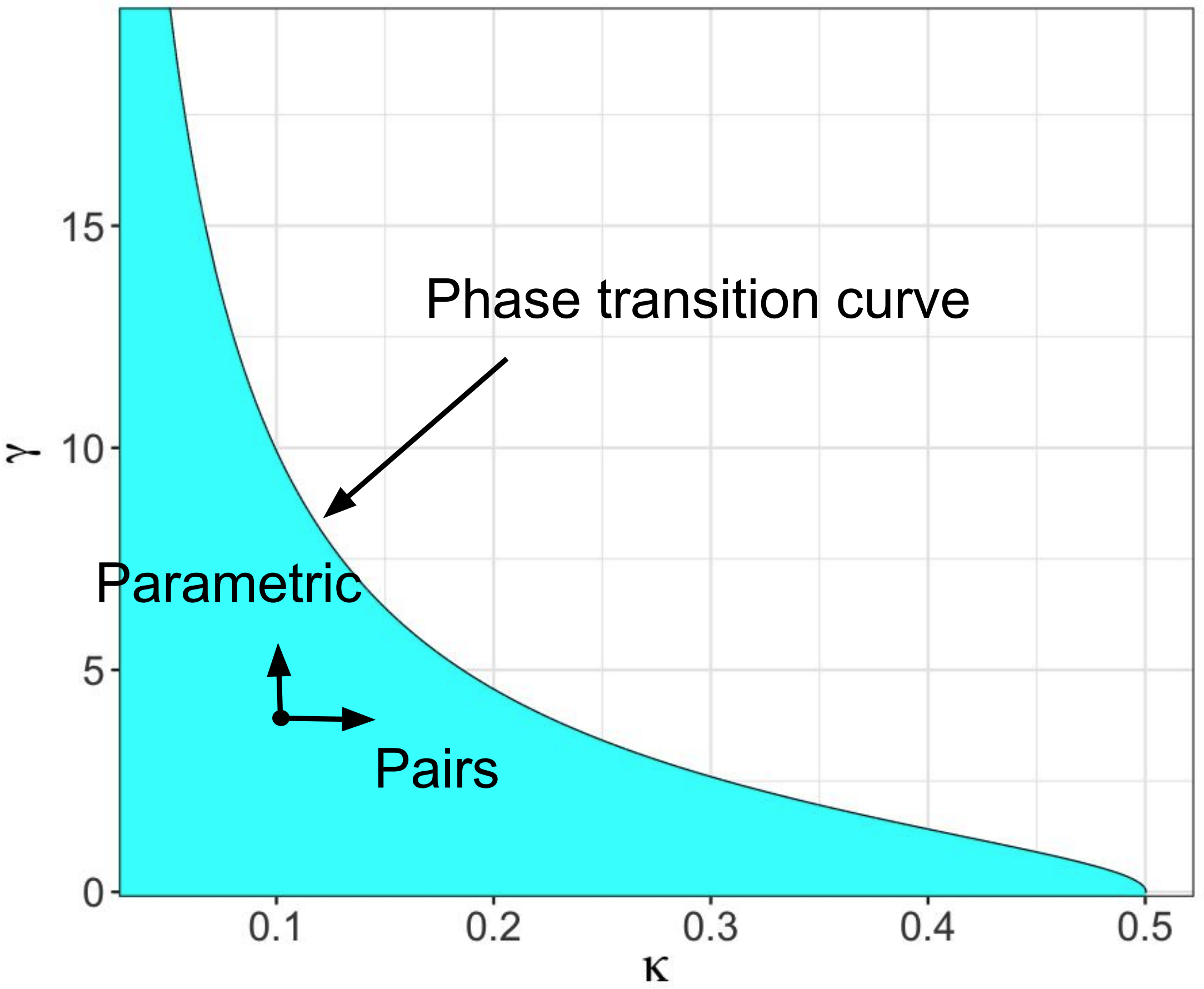}
    \caption{According to the high-dimensional theory (Section \ref{sec:high_dim_theory}), the asymptotic distribution of the MLE depends on the problem dimension $\kappa$ and the signal strength $\gamma$. The pairs bootstrap over-estimates $\kappa$ whereas the parametric bootstrap over-estimates $\gamma$. Therefore, both methods lead to incorrect estimates of the MLE distribution. The blue region shows pairs of values of $(\kappa, \gamma)$ where the MLE exists.}
    \label{fig:why_boot_fail}
\end{figure}

In addition to over-estimating the bias and standard error, another
problem of using the bootstrap is that when working with bootstrap
samples, the MLE may cease to exist. We can explain this issue via the
phase transition: for every ratio $\kappa$ and intercept $\beta_0$,
there exists an asymptotic threshold $\gamma(\kappa, \beta_0)$ such
that the MLE does not exist once the signal strength
$\gamma>\gamma(\kappa, \beta_0)$. Similarly, for every $\gamma$ and
$\beta_0$, there exists a threshold $\kappa(\gamma, \beta_0)$ such
that the MLE does not exist once $\kappa>\kappa(\gamma,
\beta_0)$. Because the pairs bootstrap over-estimates $\kappa$ while
the parametric bootstrap over-estimates $\gamma$, the bootstrap MLE may
not exist if either $\kappa$ or $\gamma$ exceeds the phase transition
threshold. Figure \ref{fig:why_boot_fail} provides a visual illustration of these points. 

\section{A resized bootstrap}\label{sec:method}

We proposing constructing parametric bootstrap samples from
$F_{\beta_\star}$, where $\beta_\star$ is obtained by shrinking the MLE
towards zero. We would like $\beta_\star$ to obey
$\var(X_{\text{new}}^\top \beta_\star) =\gamma^2 = \var(X^\top \beta)$
as to preserve the signal-to-noise ratio.  We set out to estimate
$\gamma$ in Section \ref{sec:signal_strength} since $\gamma$ is
unobserved. Upon obtaining $\beta_\star$, we follow the standard
parametric bootstrap procedure to generate bootstrap samples. That is
to say, the $b$th bootstrap sample consists of $({x}_i, Y_i^b)$,
$i = 1, \ldots, n$, where ${x}_i$ is the vector of features for the
$i$th sample and $Y^b_i$ is sampled from our GLM with features $x_i$
and coefficients ${\beta}_{\star}$. We then compute the bootstrap MLE
$\hat{{\beta}}^b\in\R^p$ by fitting the GLM using pairs
$(x_i, Y_i^b)$. Repeating this process $B$ times yields $B$ bootstrap
MLEs. We then infer the inflation and std.~dev.\ of the MLE from the
bootstrap MLE.

We summarize the procedure in Algorithm \ref{alg:resized_bootstrap} and
discuss how to compute confidence intervals using the bootstrap MLE in
Section \ref{section:boot_ci}. We evaluate our method through
simulated examples in Section \ref{sec:empirical_known_param}.

\subsection{Estimating the signal strength}\label{sec:signal_strength}

Since we would like to have
$\var({X_{\text{new}}}^\top {\beta}_{\star}) =\gamma^2$, we discuss how to
estimate $\gamma$ from observed data (see Algorithm
\ref{alg:estimate_signal_strength} for a summary). We begin by
reviewing the existing ProbeFrontier method, which applies to Gaussian
covariates, and then introduce a new approach applicable to general
covariate distributions.

The ProbeFrontier method \cite{sur2019p2} estimates $\gamma$ by using the phase
transition curve $\kappa(\beta_0, \gamma)$: if the intercept equals
$\beta_0$ and the signal strength equals $\gamma$, then the MLE does
not exist almost surely (asymptotically) if
$\kappa > \kappa(\beta_0, \gamma)$; that is, the cases and controls
can be perfectly separated by a hyperplane (see Section
\ref{sec:why_bootstrap_fails}). The ProbeFrontier method identifies
the threshold $\hat{\kappa}_s$ at which the MLE ceases to exist by
subsampling observations. It then estimates $\hat{\gamma}$ in such a
way that $\kappa(\beta_0, \hat{\gamma}) = \hat{\kappa}_s$
holds. While the ProbeFrontier method accurately estimates $\gamma$
when the covariates are Gaussian, it does not apply here because the
phase transition curve actually depends on the covariate distribution. For example, if the covariates are from a multivariate
$t$-distribution, then the phase transition curve depends on the degrees of freedom of the $t$-distribution
\cite{TangYe2020}. 
 
\begin{figure}[ht]
    \centering
    \includegraphics[width = 0.6\textwidth]{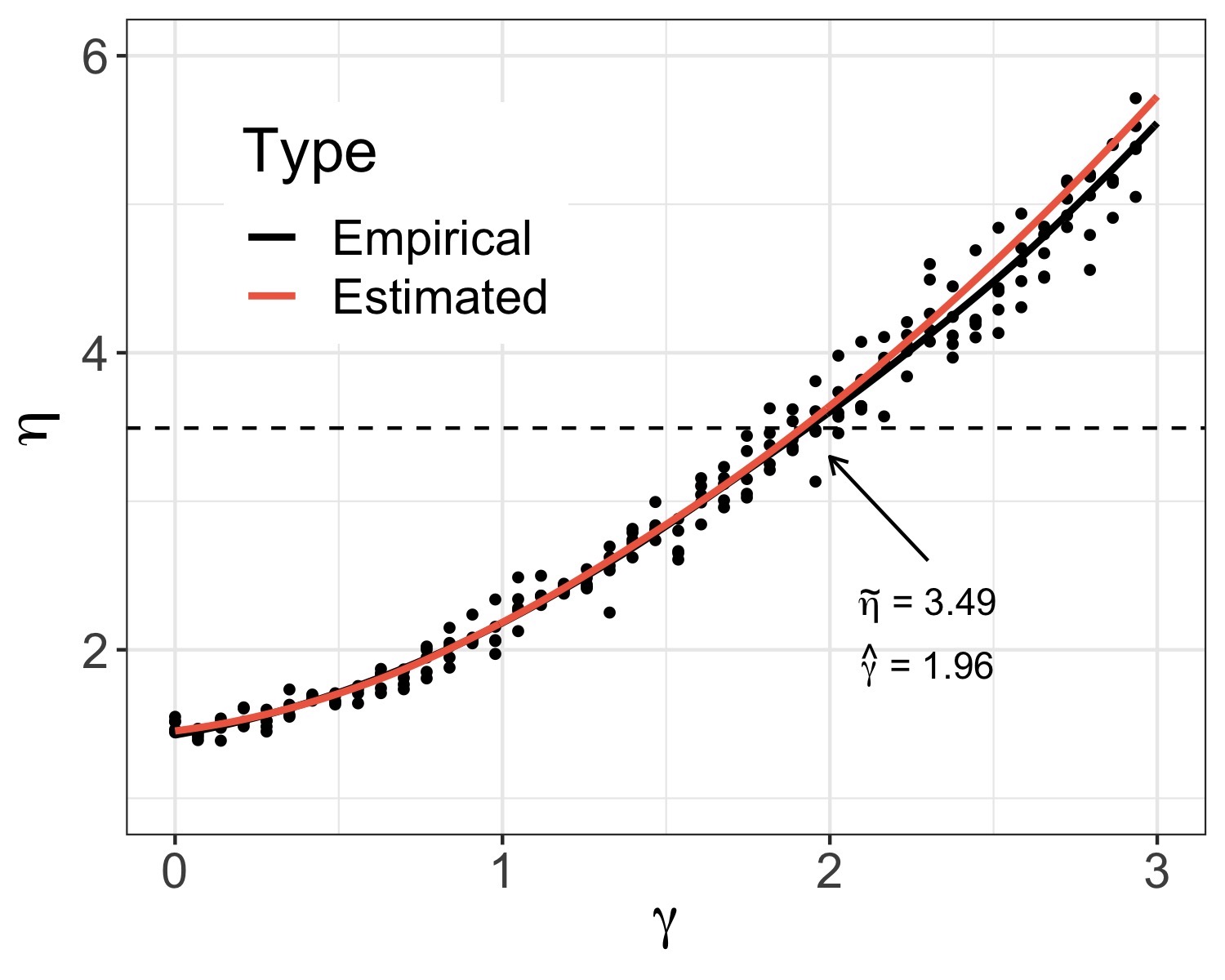}
    \caption{An illustration of using
      $\eta = {\sd}(\xnewt \hat{\beta})$ to estimate the signal
      strength $\gamma = {\sd}(\xnewt \beta)$. The orange curve shows
      $\eta$ versus $\gamma$. This is obtained by  generating 100 random samples for each $\gamma$, the orange curve being 
      the smoothed LOESS fit.
      The black curve shows an estimated curve
      using one dataset only; it is a smoothed version of
      $\hat{\eta}(\gamma)$ (black points). The dashed line shows
      $\tilde{\eta}$, and the estimated $\hat{\gamma} = 1.96$
      approximates $\gamma = 2$ well. Here, we sample covariates from
      a multivariate $t$-distribution and responses from a logistic
      model. The coefficients ${\beta}$ are sampled once and then
      re-scaled to achieve a value of $\gamma$ shown on the
      $x$-axis.  }
    \label{fig:curve1}
\end{figure}

As an alternative, we estimate $\gamma$ by using a one-to-one relation between
\begin{equation}\label{eq:defn_eta}
\gamma^2 = \var(\xnewt {\beta}) \quad \text{and}\quad \eta^2 = \var(\xnewt \hat{\beta}). 
\end{equation}
The orange curve in Figure \ref{fig:curve1} plots $\eta$ as $\gamma$
varies, and we observe that $\eta(\gamma)$ increases monotonically when
$\gamma$ increases. (Once again, this is because both the bias and the variance
of the MLE increase as $\gamma$ increases
\cite[Figure~7]{sur2019p2}.) Since the MLE does not exist when
$\gamma$ exceeds the phase transition threshold,\footnote{The phase
  transition threshold $\gamma_s$ satisfies
  $\kappa(\beta_0, \gamma_s) = \kappa$.} we expect that
$\eta$ would increase to infinity as $\gamma$ approaches the
threshold.

The one-to-one relation between $\gamma$ and $\eta(\gamma)$ suggests that, if $\var(X^\top \beta_\star) \cong \var(X^\top \beta)$, then $\var(X^\top \hat{\beta}_\star) \cong \var(X^\top \hat{\beta})$, where $\hat{\beta}_\star$ denotes the MLE when the true coefficient is $\beta_\star$. Thus, we estimate $\gamma^2$ by $\var(X^\top \beta_\star)$, where $\beta_\star$ obeys
\begin{equation}\label{eq:condition_2}
     \var(\xnewt \hat{\beta}_\star) = \eta^2. 
\end{equation}
In this paper, we set $\beta_\star$ to be a rescaled version of the MLE,
i.e., $\beta_\star = s \times \hat{\beta}$. Because the MLE is biased
upwards in absolute magnitude, the rescaling factor $s$ is less than
one and shrinks the MLE towards zero.

Although we cannot compute $\eta$ directly because it is evaluated at
a new observation $\xnew$, we estimate $\eta$ by using the SLOE
estimator introduced in \cite{yadlowsky2021sloe}. We briefly describe
SLOE here, and defer detailed formulae to Appendix~\ref{app:sloe}. The
SLOE estimator proceeds in two steps. First, it approximates
$\var(\xnewt \hat{\beta})$ by the variance of
$x_i^\top \hat{\beta}_{(i)}$ where $\hat{\beta}_{(i)}$ is the
leave-$i$th-observation-out MLE. Second, instead of re-evaluating
$\hat{\beta}_{(i)}$ for each observation, SLOE uses the first-order
approximation of the score equation to approximate $\hat{\beta}_{(i)}$
from the MLE. The theory is this: \cite{yadlowsky2021sloe} proves that
the SLOE estimator is consistent in logistic regression models with
Gaussian covariates. Furthermore, we expect that SLOE yields reliable
estimates for a broad class of covariates, for which the Euclidean
norm $\|X\|$ is concentrated and the Hessian at the MLE is positive
definite.

Now that we are able to approximate $\eta(\gamma)$ at a given
$\gamma$, we estimate $\eta=\var(X^\top \hat{\beta})^{1/2}$ and denote
it as $\tilde{\eta}$. Next, we estimate the curve ${\eta}(t)$ at a
sequence of signal strengths $t$, from which we estimate $\gamma$ by
setting $\hat{\gamma}$ such that
$\tilde{\eta} = \hat{\eta}(\hat{\gamma})$. To implement this, we pick
a sequence of scaling factors $\{0 = s_1,\ldots, s_I = 1\}$. At each
$s_i$, we set the coefficients to be
${\beta}^{s_i}= s_i \times {\hat{\beta}}$ and the signal strength
corresponding to $s_i$ as
$\gamma(s_i) = \sd(\boldsymbol{X} {\beta}^{s_i})$, where
$\boldsymbol{X}$ refers to the observed covariate matrix. We use
${\beta}^{s_i}$ as the true coefficient to generate new responses (as
in a parametric bootstrap) and then use this sample to obtain one
estimate of $\hat{\eta}(\gamma(s_i))$. Repeating the process $J$ times
yields $J$ estimates $\hat{\eta}_j(\gamma(s_i))$ for every $s_i$. We next fit 
a smoothed curve $\hat{\eta}(\gamma(s_i))$ through the 
points $\hat{\eta}_j(\gamma(s_i))$, $i=1,\ldots, I$, $j=1,\ldots,
J$. Finally, we set $\hat{\gamma}$ such that
$\hat{\eta}(\hat{\gamma}) = \tilde{\eta}$.

We demonstrate our method in Figure~\ref{fig:curve1}, which shows
$\hat{\eta}(\gamma(s_i))$ estimated from a single dataset. The
estimated curve offers an excellent fit across all values of
$\gamma$. In this example, the estimated $\tilde{\eta} = 3.49$
(dashed horizontal line), and this corresponds to
$\hat{\gamma} = 1.96$ on the black curve. This estimate is close to
the actual signal strength set to $\gamma = 2$.

\begin{algorithm}[h!]
\caption{Estimating signal strength}\label{alg:estimate_signal_strength}
\KwIn{Observed data $({x}_i, y_i)$, $1\leq i\leq n$, and a GLM formula. }
Estimate $\tilde{\eta} = \var(X^\top \hat{\beta})$ via leave-one-out techniques\;
Pick a sequence $\{0 = s_1, \ldots, s_I = 1\}$\;
        \For{$i=1,\ldots I$}{
        Set ${\beta}^{s_i} = s_i\times \hat{\beta}$  and $\gamma_i = \mathrm{sd}(\bvec{X}\beta^{s_i})$\;
        \For{$j=1,\ldots, J$}{
        Simulate $Y_i^j$ given ${x}_i$ using ${\beta}^{s_i} $ as model coefficients\;
        Fit a GLM for $({x}_i, Y_i^j)$ to estimate $\hat{\eta}_{j}(\gamma(s_i))$\; 
        }
        }
Fit a smooth curve $\hat{\eta}(\gamma)$\;
Estimate $\hat{\gamma}$ by solving $\hat{\eta}(\hat \gamma) = \tilde{\eta}$\;
\KwOut{Estimated $\hat{\gamma}$.} 
\end{algorithm}
 
\subsection{Constructing confidence intervals}\label{section:boot_ci}
We consider two ways of computing confidence intervals (CI) from
bootstrapped MLEs: first, assuming that the MLE is approximately
Gaussian, i.e.,
\begin{equation}\label{eq:assumption}
    \frac{\hat{\beta}_j - {\alpha}_j \beta_j}{\sigma_j} \approx \N(0,1), 
\end{equation}
where $\alpha_j$ and $\sigma_j$ denote the bias and standard
deviation, inverting Eqn.~\eqref{eq:assumption} yields the following
$(1-q)$ CI for $\beta_j$: 
\begin{equation}\label{eq:normal_ci}
\left[\frac{1}{\hat{\alpha}_j}\left(\hat{\beta}_j - z_{1-q/2}\,\hat{\sigma}_j\right), 
\frac{1}{\hat{\alpha}_j}\left(\hat{\beta}_j - z_{q/2}\,\hat{\sigma}_j\right)\right].
\end{equation}
Here, $z_q$ is the quantile of a standard Gaussian, while
$\hat{\alpha}_j$ and $\hat{\sigma}_j$ refer to estimates of $\alpha_j$ and $\sigma_j$. 

When the normal approximation is inadequate, we use the approximation 
\begin{equation}\label{eq:boot_assumption}
  \frac{\hat{\beta}_j - {\alpha}_j \beta_j}{\sigma_j} \stackrel{d}{\approx} \frac{\hat{\beta}^b_j - {\alpha}_j \beta_{\star, j}}{{\sigma}_j},
\end{equation} 
where the right-hand side refers to the distribution of $\hat{\beta}_j^b$ conditional on the observed covariates. After plugging in the estimated $\hat{\alpha}_j$ and $\hat{\sigma}_j$, we obtain a $(1-q)$ CI as 
\begin{equation}\label{eq:boot_ci}
\left[\frac{1}{\hat{\alpha}_j}\left(\hat{\beta}_j - t_j^b[1-q/2]\,\hat{\sigma}_j\right), 
\frac{1}{\hat{\alpha}_j}\left(\hat{\beta}_j - t_j^b[q/2]\,\hat{\sigma}_j\right)\right],
\end{equation}
where $t_j^b[q]$ denotes the quantile of the right-hand side of \eqref{eq:boot_assumption}. We refer to the confidence interval in \eqref{eq:boot_ci} as the ``bootstrap-$t$'' confidence interval, and examine the approximation \eqref{eq:boot_assumption} in Section \ref{sec:small_known_parameter}. 

Finally, we describe how to estimate the bias $\alpha_j$ and the standard deviation $\sigma_j$. To estimate $\sigma_j$, we use the standard deviation of the bootstrap MLE, i.e., 
\begin{equation}\label{eq:estimate_variance}
    \hat{\sigma}_j^2 = \frac{1}{B-1}\sum_{b=1}^B (\hat{\beta}_j^b -  \bar{\beta}_j)^2,\quad \text{where}\quad \bar{\beta}_j = \frac{1}{B}\sum_{b=1}^B \hat{\beta}_j^b.
\end{equation}
We estimate $\alpha_j$ by weighted regression: that is, we regress
$\bar{{\beta}}^b$ onto $\beta_{\star}$ by assigning to each MLE coordinate
a weight inversely proportional to its estimated variance
$\hat{\sigma}_j^2$. We assume a common bias factor because all the
${\alpha}_j$'s are equal when the covariates are multivariate
Gaussian. In practice, we can plot $\bar{\beta}_j^b$ versus
$\beta_{\star, j}$: if bias factors are all equal, then the points should
align on a line, which we observe in all our simulations
(Section~\ref{sec:empirical_known_param}). 

\begin{algorithm}[h!]
\caption{Resized bootstrap procedure}\label{alg:resized_bootstrap}
\KwIn{Observed data $({x}_i, y_i)$, $1\leq i\leq n$, and a GLM formula.}
Compute resized coefficients ${\beta}_{\star}$\;
\For{$b = 1, \ldots, B$}{
        Simulate $Y_i^b$ given ${x}_i$ using ${\beta}_{\star}$ as model coefficients\;
         Fit a GLM for $({x}_i, Y_i^b)$ to obtain the bootstrap MLE $\hat{{\beta}}^b$\;
         }
Estimate the standard deviation of the MLE $\hat{\sigma}_j$ (See Eqn.~\eqref{eq:estimate_variance})\;
Estimate a common factor $\hat{\alpha}$ by regressing $\bar{{\beta}}$ onto ${\beta}_{\star}$ with weights proportional to $1/\hat{\sigma}_j^2$\;
\KwOut{$\hat{\alpha}$ and $\hat{\sigma}_j$} 
\end{algorithm}

\subsection{When is the resized bootstrap adequate?}\label{section:boot_assumption}
When the covariates are multivariate Gaussian, \cite{zhao2020} observed that while Eqn.~\eqref{eq:gaussian_asym_theory} is accurate when $\beta_j$ is moderately large, \footnote{Here, we assume the covariates $X_j$ are standardized to have zero mean and unit variance.} the std.~dev.\ of $\hat{\beta}_j$ increases as the absolute magnitude of $\beta_j$ increases. This result implies that the resized coefficient $\beta_{\star,j}$ should be close to $\beta_j$ in order to correctly estimate the MLE distribution. However, the resized coefficients only satisfy
$\var(X^\top {\beta}_\star) = \gamma^2$, and yet $\beta_{\star_j} \neq \beta_j$ in general. Therefore, we expect that the CIs to be approximately correct when $\beta_j$ is moderately large, but inaccurate when $\beta_j$ is large. We explore the performance of our method when the model
coefficients are large in
Appendix~\ref{appendix:resizedboot_assumption}. While we expect that
correct inference can be obtained by shrinking the large and small
coefficients separately, we leave this study for
future research.

\section{Numerical studies} \label{sec:empirical_known_param}

We now study the accuracy of the proposed resized bootstrap method by simulating GLMs with non-Gaussian covariates. We consider logistic regressions in Section \ref{section:simulation_result}-\ref{sec:small_known_parameter} and Poisson regression in Section \ref{sec:poisson}. We report results of other examples in Appendix \ref{appendix:additional_simulation}. \footnote{We refer readers to \url{https://github.com/zq00/glmboot} for the \texttt{R} code to implement simulations in this article. }

\subsection{Simulation design}\label{section:simulation_design}
First, we set $n = 4000$ and $p = 400$ ($\kappa = p/n = 0.1$). Second,
we consider two cases of covariate distributions:
\begin{enumerate}
\item The covariates follow a multivariate $t$-distribution (MVT) with
  $\nu = 8$ degrees of freedom whose covariance matrix $\Sigma$ is a
circulant matrix equal to $\Sigma_{ij} = 0.5^{\min(|i - j|, p+1 - |i-j|)}$. This
  structure implies that the variance of a predictor conditional on
  the others is the same regardless of the predictor. In turn, HDT
  then predicts that all the MLE coefficients have equal standard
  deviation.
\item The covariates follow a modified ARCH model
  ${X} = \zeta {\varepsilon}$, where $\zeta$ is the inverse of a
 $\chi$ variable with $\nu = 8$ degrees of freedom \footnote{A $\chi$ variable is distributed as the square root of a chi-squared variable} and
  ${\varepsilon}$ is from an Autoregressive Conditional
  Heteroskedasticity (ARCH) model (see
  \cite[Section~5.4]{shumway2017book} for a definition of ARCH
  models). Here, starting with $X_0 \sim \N(0, \alpha_0/(1-\alpha_1))$, we
  sequentially sample variables so that $X_j= \sigma_j \varepsilon_j$,
  where $\sigma_j^2 = \alpha_0 + \alpha_1 X_{j-1}^2$ and
  $\varepsilon_j\sim\N(0,1)$. We work with $\alpha_0 = 0.6$ and
  $\alpha_1 =0.4$.  Although uncorrelated, the covariates are not
  independent of each other. 
\end{enumerate}

After sampling the covariates, we sample responses from a logistic model. We sample  model coefficients by first picking 50 non-null variables; then, we sample the magnitude of the non-null coefficients from an equal mixture of $\N(5,1)$ and $\N(-5,1)$. This signal strength ensures that the MLE exists. At the same time, the signal strength is strong enough such that we can tell a large proportion the the non-null variables apart from the nulls. For instance, when $\beta_j=4.78$ as in the example in Section \ref{sec:intro_example} , over 90\% of the 95\% CI excludes 0, and approximately 90\% of the non-null coefficients from the mixture distribution satisfy this property.  

\subsection{Results}\label{section:simulation_result}
We report below the estimated bias and standard deviation of the MLE
as well as the coverage proportions. We also examine the MLE
distribution and the assumption that the bias factors $\alpha_j$ are
all equal. Without specifying, we consider covariates from a multivariate $t$-distribution.

\subsubsection{Estimated Bias and Variance}
\begin{figure}[h!]
    \centering
    \includegraphics[width = 0.45\textwidth]{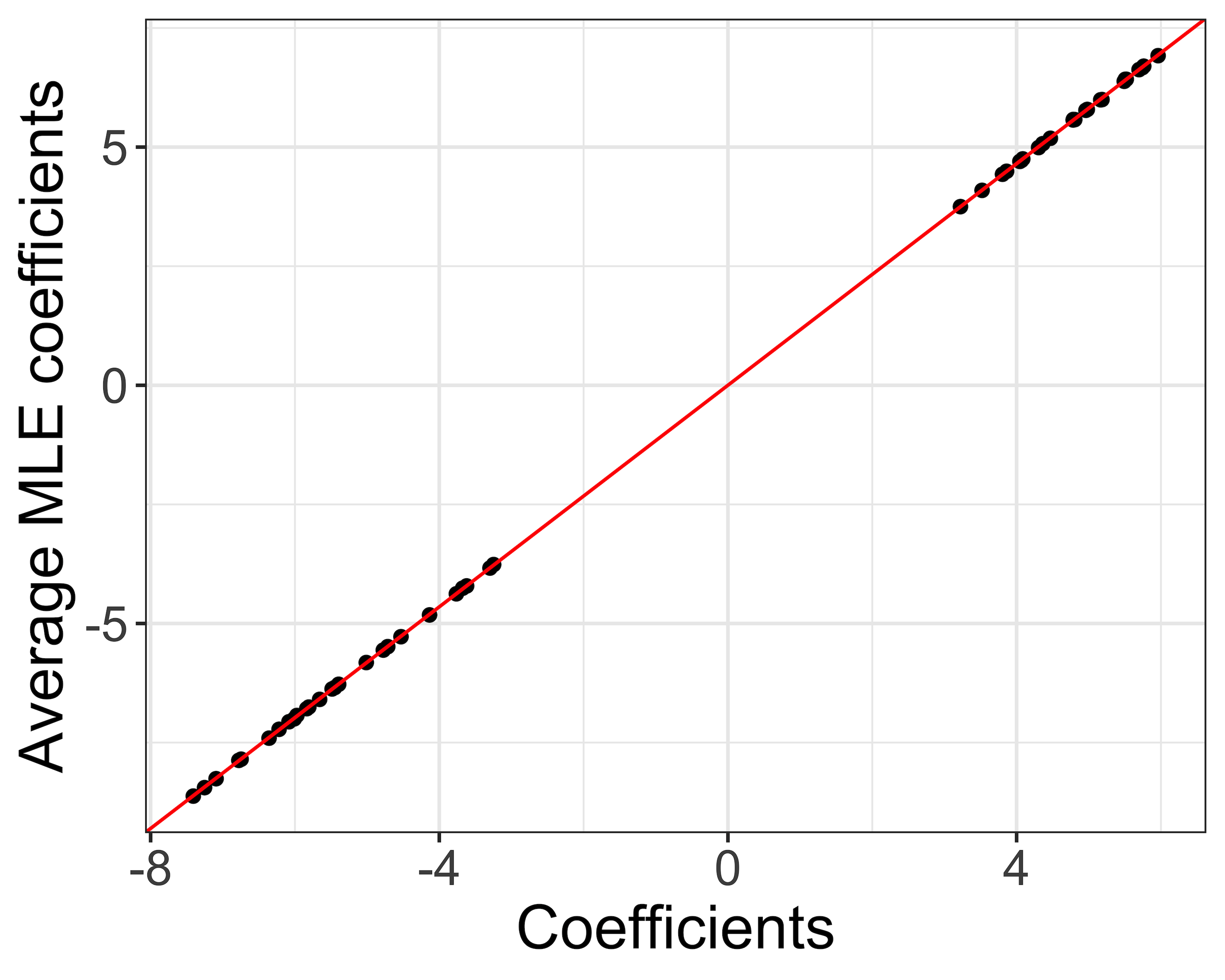}
    \caption{Average MLE versus model
      coefficients for the non-null variables. The $x$-axis shows the magnitude of each non-null coefficient and the $y$-axis shows the average MLE over 10,000 repetitions. The red line has zero intercept and slope equal to 1.16. }
    \label{fig:bias_resized_boot}
\end{figure}

From Section \ref{sec:high_dim_theory} we know that the MLE is just
too sure in the sense that the estimated magnitude is biased
upwards. As an illustration, Figure~\ref{fig:bias_resized_boot} plots
the average MLE versus the model coefficients when the covariates are
from (modified) ARCH model above. Since the scatterplot lies near
a line, we can see that the $\alpha_j$'s do not seem to much depend on
the magnitude of the coefficients; additionally, the plot confirms the
bias of the MLE since the line has a slope greater than 1. For
information, we get a very similar plot for the multivariate $t$-covariates.  

We now examine the accuracy of the estimated bias using existing high-dimensional theory and the resized bootstrap (recall that both estimate a common bias factor). Table \ref{tab:bias_variance} reports the estimated bias and variance of a single null and a single non-null variable. As observed in Section \ref{sec:intro_example}, HDT captures the bias, and Table \ref{tab:bias_variance} shows that the resized bootstrap estimate is also reasonably accurate. As to the standard deviation, while both methods slightly underestimate the std.~dev., the resized bootstrap is more accurate and its relative error is less than 3\%. In particular, the resized bootstrap captures the increased std.~dev.\ of the MLE of non-null variables in comparison to null variables. In contrast, classical calculations based on the Fisher information significantly underestimate the std.~dev.. Since the resized bootstrap yields a more accurate std.~dev., we would expect enhanced CIs. 
\begin{table}[h!]
    \centering
    \begin{tabular}{|l| c c c | c c c c|}
    \hline
     & \multicolumn{3}{|c | }{Bias}  & 
    \multicolumn{4}{c | }{Standard Deviation} \\
    \hline
     &  High-dim  &  Resized  &  Empirical & Classical & High-dim & Resized   & Empirical \\
     &  Theory  & Bootstrap&   Bias  & Theory & Theory &  Bootstrap & Std.~dev. \\
         \hline
    $\beta = 0$ & - & - & -  &  1.232 & 1.259 & 1.316  & 1.327 \\
    $\beta = 5.519$ & 1.151 & 1.159  & 1.160 & 1.244  & 1.259   & 1.327 & 1.337 \\
    \hline
    \end{tabular}
   \caption{Estimated bias and std.~dev.\ of the MLE. The correct values (empirical bias and std.~dev.) have been obtained from 10,000 repetitions. The std.~dev.\ from classical theory is calculated by the \texttt{glm} function in \texttt{R} and averaged over 10,000 repetitions. The resized bootstrap estimates are computed by taking an average over 1000 repetitions and uses an estimated signal strength $\gamma$. }
        \label{tab:bias_variance}
\end{table}

\subsubsection{Coverage Proportion}

\begin{figure}[ht]
    \centering
    \includegraphics[width = 0.4\textwidth]{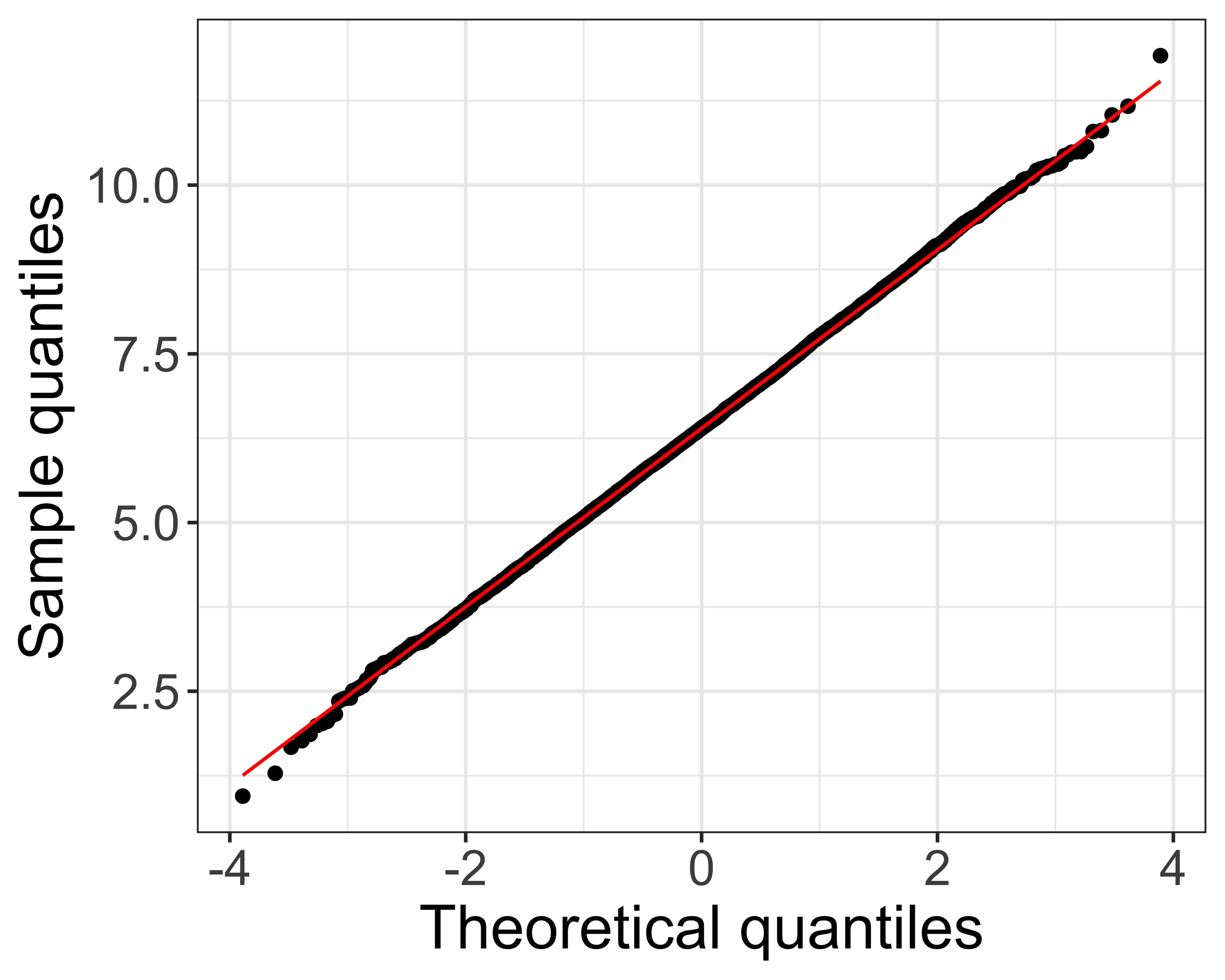}
    \hspace{1cm}
    \includegraphics[width = 0.4\textwidth]{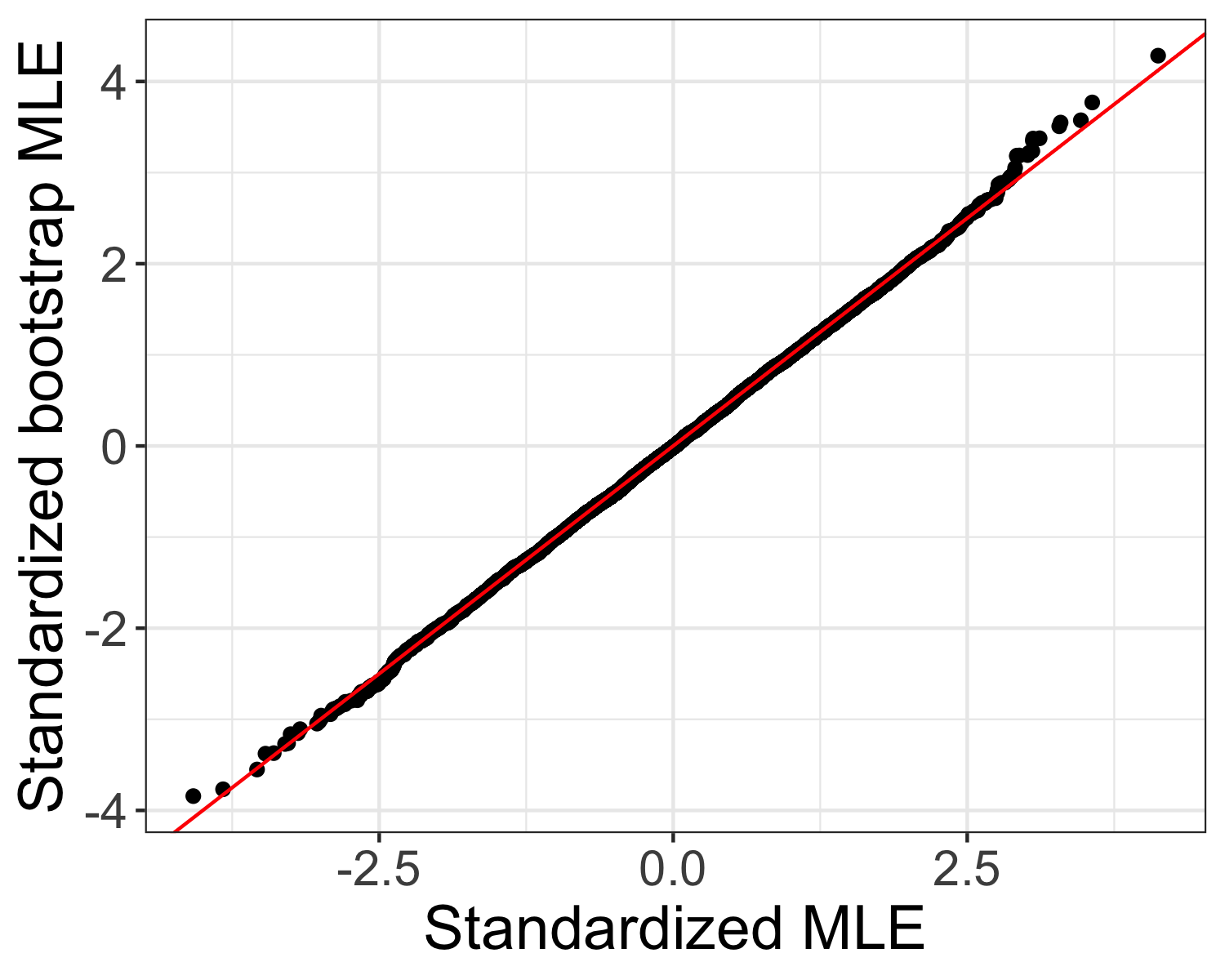}
    \caption{(Left) Normal Q-Q plot of the MLE. (Right) Q-Q plot of the standardized bootstrap MLE (in one simulated example) versus the standardized MLE. In this example, the covariates are sampled from a multivariate $t$-distribution.}
    \label{fig:resized_boot_qq}
\end{figure}

Section \ref{section:boot_ci} introduced two types of CIs, based on
the assumptions that the MLE is approximately Gaussian
(Eqn.~\eqref{eq:assumption}) or that the standardized bootstrap MLE
approximates the distribution of the standardized MLE
(Eqn.~\eqref{eq:boot_assumption}). Before evaluating accuracy, we
examine these assumptions by showing a normal Q-Q plot of the MLE
(Figure~\ref{fig:resized_boot_qq}, Left) and a Q-Q plot of the
standardized bootstrap MLE versus the standardized MLE
(Figure~\ref{fig:resized_boot_qq}, Right). Here, we standardize the
bootstrap MLE by the estimated bias and estimated std.~dev.\ and the MLE by the correct bias
and std.~dev.. Along the points align on the 45 degree line in both plots, we conclude that both assumptions are
reasonable and, therefore, expect that both CIs would perform well.

Denote the confidence interval for $\beta_j$ in the $i$th simulation as $\ci_{i,j}$, and define the proportion of times a single variable $\beta_j$ is covered (Table \ref{tab:empirical_cov_single}) as 
\begin{equation}\label{eq:qj}
   q_j := \frac{1}{N}\sum_{i=1}^N \ind\{\beta_j\in \ci_{i,j}\}. 
\end{equation}
Define the coverage proportion of {\em all of the variables in the $i$th experiment only} (Table \ref{tab:empirical_cov_bulk}) as 
\begin{equation}\label{eq:qbar}
    \bar{q}_i = \frac{1}{p}\sum_{j=1}^p \ind\{\beta_j\in \ci_{i,j}\}
  \end{equation}
We report both coverage of a single coefficient $q_j$ and the proportion of variables covered in a {\em single-shot experiment} $\bar{q} = \frac{1}{N}\sum_{i=1}^N \bar{q}_i$ in Tables \ref{tab:empirical_cov_single} and \ref{tab:empirical_cov_bulk}. Both the Gaussian approximation (Boot-$g$) and bootstrap MLE distribution (Boot-$t$) are used to compute the CIs. The two CIs not only differ in their formulae, but also in the number of bootstrap samples: we use $B=10,000$ bootstrap samples to compute the boot-$t$ CI, but only $B=100$ bootstrap samples to compute the boot-$g$ CI. This is because boot-$g$ CI requires only estimates of the bias and variance, while boot-$t$ CI requires an estimate of the entire distribution.

\begin{table}[ht]
    \centering
    \begin{tabular}{|c|c c | c c | c c  c c|}
        \hline
    & \multicolumn{2}{|c | }{Theoretical CI}  & 
    \multicolumn{2}{c | }{Standard Bootstrap} &
    \multicolumn{4}{c|}{Resized Bootstrap}\\
      Nominal   &  & & & & \multicolumn{2}{c}{Known $\gamma$}  & \multicolumn{2}{c|}{Estimated $\gamma$} \\
      coverage & Classical & High-Dim & Parametric & Pairs & 
      Boot-$g$  & Boot-$t$ & Boot-$g$  & Boot-$t$ \\
      \hline
\multirow{ 2}{*}{95}  & 87.3 & 93.5  & 71.1 & 76.3 & 93.6 & 93.9 & 94.2 & 94.4 \\
& (0.3) & (0.3)  & (1.6)& (1.3) & (0.7) & (0.7) & (0.8) & (0.8)\\

 \multirow{ 2}{*}{90}  & 79.4 & 87.9 & 61.2 & 66.6 & 88.5 & 88.7 & 88.6 & 89.1 \\
 &  (0.30) & (0.3) & (1.7) & (1.4) & (1.0) & (1.0) & (1.1) & (1.1)\\
 
\multirow{ 2}{*}{80}  & 67.4 & 77.2  & 46.8 & 52.7 &  79.5 & 79.6 & 80.8 & 80.0 \\
&(0.5) & (0.4)  & (1.7) & (1.5) & (1.2) & (1.2) & (1.3) & (1.4)\\
\hline
    \end{tabular}
  \caption{Coverage proportion of a single variable ($q_j$ in Eqn.~\eqref{eq:qj}) with standard deviation between parentheses. This example uses multivariate-$t$ covariates.}
    \label{tab:empirical_cov_single}
\end{table}

\begin{table}[ht]
    \centering
     \begin{tabular}{|c|c c | c c | c c  c c|}
        \hline
    & \multicolumn{2}{|c | }{Theoretical CI}  &
     \multicolumn{2}{c | }{Standard Bootstrap}  &\multicolumn{4}{c|}{Resized Bootstrap}\\
      Nominal   &  & & & & \multicolumn{2}{c}{Known $\gamma$}  & \multicolumn{2}{c|}{Estimated $\gamma$} \\
      coverage & Classical & High-Dim & Parametric & Pairs & Boot-$g$  & Boot-$t$ & Boot-$g$  & Boot-$t$ \\
      \hline
\multirow{ 2}{*}{95}  & 92.5 & 93.7 & 90.8 & 93.3 &  94.6 & 94.9 & 94.7 & 95.0 \\
& (0.02) & (0.02) & (0.06) & (0.05) & (0.04)  & (0.04) & (0.04) & (0.04)\\

 \multirow{ 2}{*}{90}  & 86.6 & 88.2& 84.5 & 87.8 & 89.5  & 89.7 & 89.7 &89.9 \\
 & (0.02)& (0.02) & (0.08) & (0.06) & (0.06)& (0.06)& (0.06) & (0.06)\\
 
\multirow{ 2}{*}{80}  & 75.7& 77.7& 73.6 & 77.5 & 79.4 & 79.5 & 79.6 & 79.7 \\
& (0.03) & (0.03) & (0.09) & (0.08) & (0.08) & (0.08) & (0.08) & (0.08)\\
\hline
    \end{tabular}
      \caption{The proportion of covered variables in a single-shot experiment ($\bar{q}$ in Eqn.\ \eqref{eq:qbar}). The standard deviation is given between parentheses. }
    \label{tab:empirical_cov_bulk}
\end{table}

While the resized bootstrap slightly undercovers a single coefficient
(Table \ref{tab:empirical_cov_single}), the relative error is within
1.5\% in all of the levels we examined. Similarly, the proportion of
variables covered in a single-shot experiment (Table
\ref{tab:empirical_cov_bulk}) is also close to the nominal coverage
and the relative error is within 1\%. In addition, boot-$g$ and
boot-$t$ CI achieve similar accuracy at every level we examined. Since
boot-$g$ CI uses a smaller sample size, we prefer boot-$g$ CI when the
Gaussian assumption holds. We can verify the
normality assumption by comparing the quantiles of bootstrap MLEs with
normal quantiles.
Table \ref{tab:empirical_cov_single} shows the coverage of a non-null
variable, and we report coverage of a null variable in
Appendix \ref{appendix:null}. Comparing the
coverage probability using the estimated signal strength
$\hat{\gamma}$ versus its true value $\gamma$ shows that the method
with estimated parameters perform as well as if we had an oracle.

As to the other methods, the HDT CIs slightly undercover since variability
is underestimated as seen earlier. Classical CIs significantly
undercover. Neither the parametric nor the pairs bootstrap provide the
correct coverage, and this is consistent with observations from Figure
\ref{fig:intro_ex}.

\subsection{Small sample sizes}\label{sec:small_known_parameter}

We study an example with a small sample size, and set $n = 400$ and $p=40$. We sample covariates independently from a Pareto distribution\footnote{The Pareto distribution is heavy-tailed and its density is $f(x) = \alpha M^\alpha  / x^{\alpha + 1}$ where $\alpha$ is the shape parameter and  $M$ is the scale parameter. We set $\alpha = 5$ and $M=1$.} and sample responses from a logistic model where half of the variables are non-nulls and sampled from an equal mixture of $\N(5, 1)$ and $\N(-5, 1)$.  

\begin{figure}[h]
    \centering
    \includegraphics[width = 0.40\textwidth]{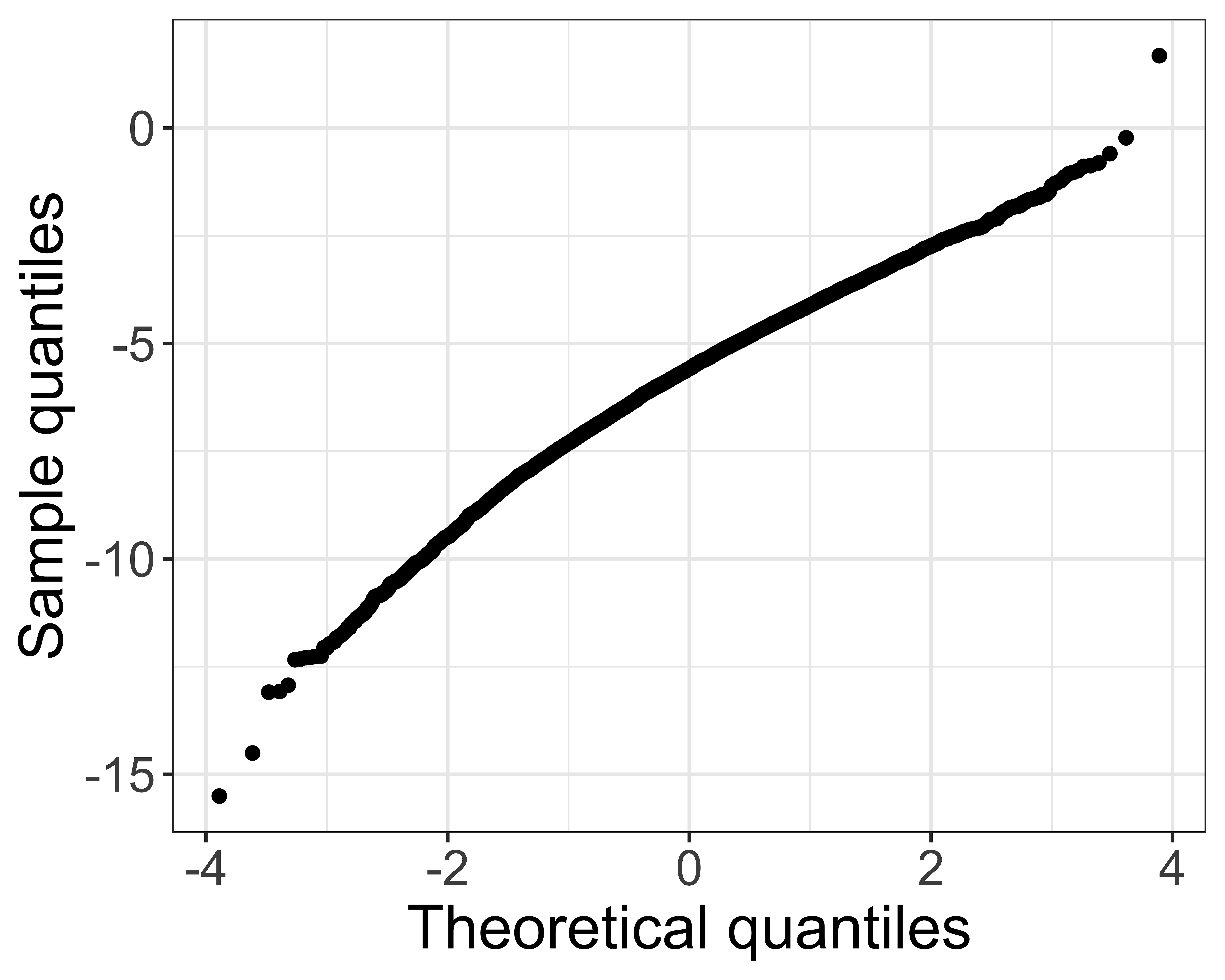}
\hspace{1cm}
    \includegraphics[width = 0.40\textwidth]{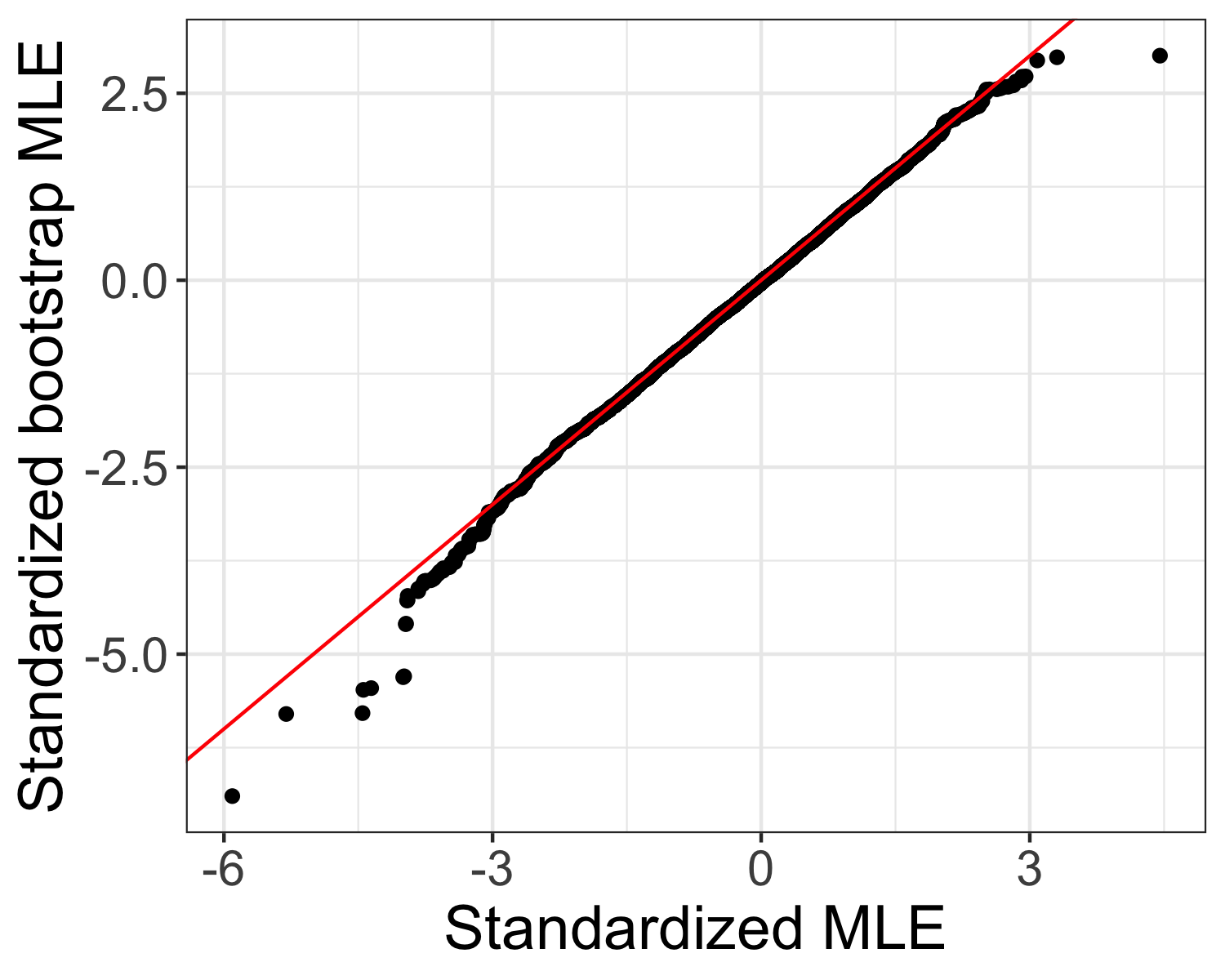}
    \caption{(Left) Normal quantile plot of a single MLE coordinate where $\beta = -4.6$ and $n=400$, $p = 40$ and the covariates are i.~i.~d.\ from a Pareto distribution. (Right) Q-Q plot of the standardized MLE versus the bootstrap MLE in a single experiment. The standardization is as in Figure \ref{fig:resized_boot_qq}. The red line is the 45 degree line. }
    \label{fig:boot_qqplot}
\end{figure}

When the covariates are i.~i.~d., the MLE may be asymptotically Gaussian, however, the normal approximation is inaccurate when $n$ is small. To see this, we can express $\hat{\beta}_j$ as 
\begin{equation}\label{eq:approximate_mle_null}
\hat{\beta}_j = \frac{\lambda}{\kappa\sqrt{n}} \sum_{i=1}^n x_{ij}(y_i - \rho'(x_{i,-j}^\top \hat{\beta}_{[-j]})) + o_P(1), 
\end{equation}
where $\hat{\beta}_{[-j]}$ refers to MLE computed when leaving out the $j$th variable and $\rho(t) = \log(1 + e^t)$
\cite[Appendix~C]{sur2019p2}. Although Eqn.\
\eqref{eq:approximate_mle_null} assumes that the $X_{ij}$ are standard
Gaussian, we expect that it holds for other i.~i.~d.\ covariates. Since
$\hat{\beta}_j$ is approximately a weighted average of the observed
$x_{i,j}$, $\hat{\beta}_j$ approaches a Gaussian random variable as
$n\to\infty$ by the central limit theorem.  Since the rate of convergence depends on the third moment of
$X_{ij}$ as a result of the Berry-Esseen theorem, we expect that the
distribution of the MLE deviates from a Gaussian distribution when $n$
is small.  Indeed, the normal
quantile plot of $\hat{\beta}_j$ (Figure~\ref{fig:boot_qqplot}, Left)
confirms that the MLE is skewed to the left and thus not Gaussian. In
comparison, the normal quantile plot when $n = 4000$ and $p=400$ (see Figure \ref{fig:normal_qq_iid_large} in Appendix \ref{appendix:qq}) indicates that the MLE is well-approximated by
a Gaussian distribution when $n$ is large.

While the MLE is not Gaussian, a Q-Q plot of the standardized
$\hat{\beta}_j^b$ versus the standardized $\hat{\beta}_j$ shows that
the bootstrap MLE approximates the sampling distribution very well
(Figure~\ref{fig:boot_qqplot}, Right). We thus expect that the resized
bootstrap provides coverage. This is confirmed in Table
\ref{tab:coverage_small}, which shows that the bootstrap CIs are
reasonably accurate for both a single variable and a single-shot
experiment across all of the confidence levels examined. As before,
the resized bootstrap using estimated parameters performs equally well
as when we know the true values.  While we do not report the coverage
proportion using bootstrap-$g$ CIs, we observe that they yield similar
accuracy as bootstrap-$t$ CI intervals. Lastly, we note that the
bootstrap MLE (standardized by estimated bias and variance) has a
lighter tail than the standardized MLE (standardized by the true bias
and variance). This explains why the resized bootstrap CI slightly
undercovers.

Our results in this example suggest that the bootstrap CIs produce
reasonable coverage when the sample size is small and the normal
approximation is far from valid.  This in contrast to methods based on
high-dimensional theory since we can see that the corresponding CI's
undercover. Again, this happens because variability is underestimated.

\begin{table}[h]
     \centering
    \begin{tabular}{|c | c c c | c c c|}
    \hline
 & \multicolumn{3}{c|}{I. Single variable} & \multicolumn{3}{c|}{II. Single Experiment}\\
 Nominal  & High-dim &  \multicolumn{2}{c|}{Resized Bootstrap} &  High-dim & \multicolumn{2}{c|}{Resized Bootstrap}  \\
 Coverage & Theory & Known-$\gamma$ & Estimated-$\gamma$ & Theory & Known-$\gamma$ & Estimated-$\gamma$ \\
\hline
95 & 88.7 (0.3) & 95.2 (0.7) & 95.7 (0.6) & 91.6 (0.1) & 94.8 (0.2) & 95.6 (0.1) \\
90 & 81.6 (0.4) & 90.0 (1.0) & 91.5 (0.9) & 86.0 (0.1) & 89.7 (0.2) & 90.9 (0.2) \\
80 & 70.0 (0.5) & 78.6 (1.3) & 81.0 (1.2) & 75.7 (0.1) & 79.2 (0.3) & 80.4 (0.2) \\
\hline
    \end{tabular}
   \caption{Coverage proportion of a single variable (Column I) and of all of the variables (Column II) in 10,000 repeated experiments with $n = 400$ and $p = 40$. We use the resized bootstrap method with known or estimated parameters.  The standard deviations are reported between parentheses.}
    \label{tab:coverage_small}
\end{table}

\subsection{A Poisson regression example}\label{sec:poisson} 
We now consider an example with Poisson regression with log link function, i.~e., $Y\,|\,X\sim \mathrm{Poisson}(\mu(X))$ and $\log(\mu(X)) = X^\top \beta$ \cite[Chapter~12]{AppliedLinearRegression2014}. We use the same simulation design as in Section \ref{section:simulation_design}, with the exception that the non-null coefficients are sampled from an equal mixture of $\N(3, 1)$ and $\N(-3, 1)$ to prevent $\mu(X)$ from being too large. We report the bias and std.~dev.\ of both a null and a non-null variable in Table \ref{tab:bias_variance_poisson}. We only use the classical theory and the resized bootstrap to estimate the std.~dev., since HDT is unavailable for Poisson regression. Table \ref{tab:bias_variance_poisson} shows that the MLE is approximately unbiased and the std.~dev.\ using the classical theory is quite accurate. The resized bootstrap also accurately estimates the std.~dev.\ of the MLE. Therefore, we expect that both approaches would produce CI with correct coverage. Indeed, the coverage proportion using both the classical theory and the resized bootstrap method are quite accurate, as the average coverage proportions are within two standard deviations away from the nominal coverage (see Table \ref{tab:coverage_poisson}). In sum, both the classical theory and the resized bootstrap yield reasonably accurate CI in case of a Poisson regression.

\begin{table}[h!]
    \centering
    \begin{tabular}{|l| c c c | c c c c|}
    \hline
     & \multicolumn{3}{|c | }{Bias}  & 
    \multicolumn{4}{c | }{Standard Deviation} \\
    \hline
     &   \multicolumn{2}{c  }{Resized Bootstrap}    &  Empirical & Classical & \multicolumn{2}{c }{Resized Bootstrap}     & Empirical \\
     &  Known-$\gamma$ & Estimated-$\gamma$ &   Bias  & Theory & Known-$\gamma$ & Estimated-$\gamma$ & Std.~dev. \\
         \hline
    $\beta = 0$ & - & - & -  &  0.272 & 0.27 & 0.27 & 0.268 \\
    $\beta = 3.24$ & 0.989 & 0.990 & 0.990 & 0.270 & 0.27 & 0.27 & 0.272\\
    \hline
    \end{tabular}
   \caption{Estimated bias and std.~dev.\ of the MLE from a Poisson regression. Please see Table \ref{tab:bias_variance} for detailed description of the table.}
        \label{tab:bias_variance_poisson}
\end{table}

\begin{table}[h!]
\centering
\begin{tabular}{|l | c | c | c  c |}
\hline
& Nominal & Classical & \multicolumn{2}{c|  }{Resized Bootstrap} \\
& Coverage & Theory & Known-$\gamma$ & Estimated-$\gamma$ \\
\hline 
\multirow{ 3}{*}{Single Null} & 95 & 95.4 (0.2) & 93.8 (0.7) & 93.6 (0.7)\\
&  90 & 90.4 (0.3)& 89.1 (0.9) & 89.3 (0.9) \\
& 80 &80.1 (0.4) & 80.7 (1.1) & 80.2 (1.2)\\
\hline
\multirow{ 3}{*}{Single Non-null} &  95 & 94.6 (0.2)& 95.1 (0.6)& 94.3 (0.7)\\
&  90 & 89.7 (0.3)& 90.5 (0.8) & 90.0 (0.9)\\
& 80 & 79.2 (0.4) & 80.0 (1.2) & 80.1 (1.17) \\
\hline
\multirow{ 3}{*}{Single Experiment} &  95 & 95.1 (0.01)& 94.7 (0.04) & 94.6 (0.04) \\
&  90 & 90.2 (0.01)& 89.6 (0.05)& 89.6 (0.05) \\
& 80 & 80.3 (0.02) & 79.7 (0.06) & 79.7 (0.07)\\
\hline
\end{tabular}
 \caption{Coverage proportion of a single null or non-null variable ($q_j$ in Eqn. \eqref{eq:qj}) and the proportion of covered variables in a single-shot experiment ($\bar{q}_i$ in Eqn. \eqref{eq:qbar}). The coverage proportion is computed using 10,000 repetitions, and 1,000 for the resized bootstrap. We use boot-$g$ CI for the resized bootstrap. The standard deviations are reported between parentheses.}
 \label{tab:coverage_poisson}
\end{table}

\section{Application to a real data set}
\label{sec:forest_covertype} 

Having observed that the resized bootstrap procedure provides more accurate inference compared to classical and high-dimensional theory, we now analyze a real data set. In this study \cite{Hong2019}, researchers aim to understand which factors are associated with restrictive spirometry pattern (RSP), which is a lung condition. In particular, they hypothesize that glomerular hyperfiltration (GHF), which assesses the kidney function, may be associated with the risk of RSP. To evaluate their hypothesis, they collected participants data from from the Korea National Health and Nutrition Examination Survey (KNHANES) from 2009-2015. They performed a logistic regression, where the response variable is RSP (defined as FVC $<$ 80\% AND FEV1/FVC $\ge$ 0.7) and the covariates include demographic variables, medical history, medications used, and a variety of health-related variables.

\begin{figure}
\centering
    \includegraphics[width = 0.75\textwidth]{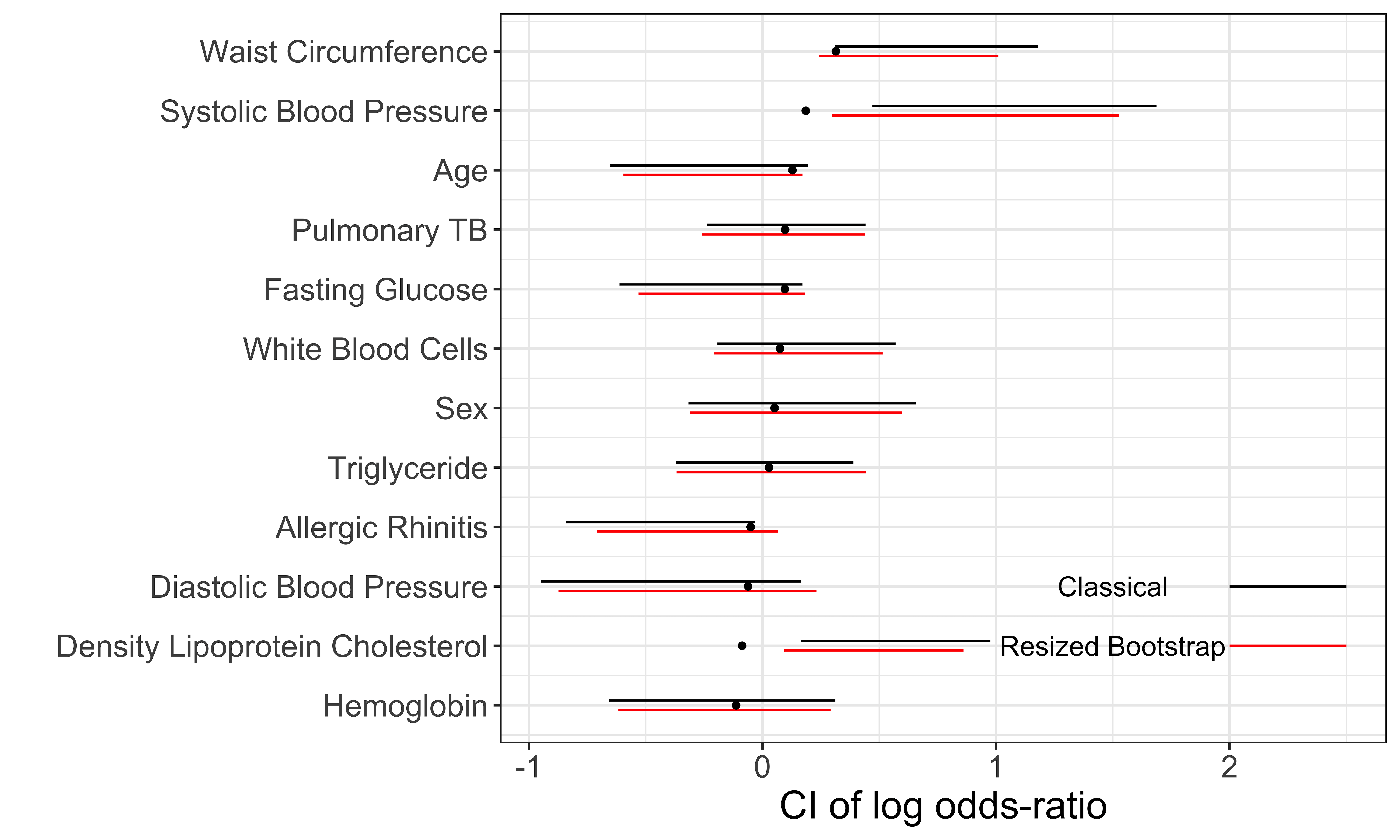}
    \caption{Confidence interval for each variable using classical theory (black) and the resized bootstrap (red). The black points indicate true model coefficients, estimated using the full data set. While we include demographic variables in the logistic model, we do not present their fitted coefficients as in Table 2 of the paper.}
    \label{fig:real_coverage}
\end{figure}

For the purpose of illustrating our approach, we fit a logistic regression using subsamples of sample size $n = 200$ and include $p=18$ covariates including the intercept ($\kappa = 18 / 200 = 0.09$).  \footnote{We only include binary variables such that both positive and negative classes occur in at least 5\% of all the samples.} We examine whether the confidence intervals of the model coefficients, i.e.\ , the log odds ratios, cover the ``true'' coefficients, which we estimate by the logistic MLE using the full data that contains about 22,000 observations. Figure \ref{fig:real_coverage} shows the CI for each covariate using classical theory (black), resized bootstrap (red), \footnote{Because the estimated $\gamma$ is random, we repeat 10 times and use the average as the estimated signal strength.} and the estimated coefficient using the full data (black points). The resized bootstrap CI is closer to zero compared to CI using the classical theory, and is slightly more accurate. For instance, the coefficient for waist circumference is covered by the red segment, but is not covered by the black segment. 

\begin{figure}[h!]
    \centering
    \includegraphics[width = 0.4\textwidth]{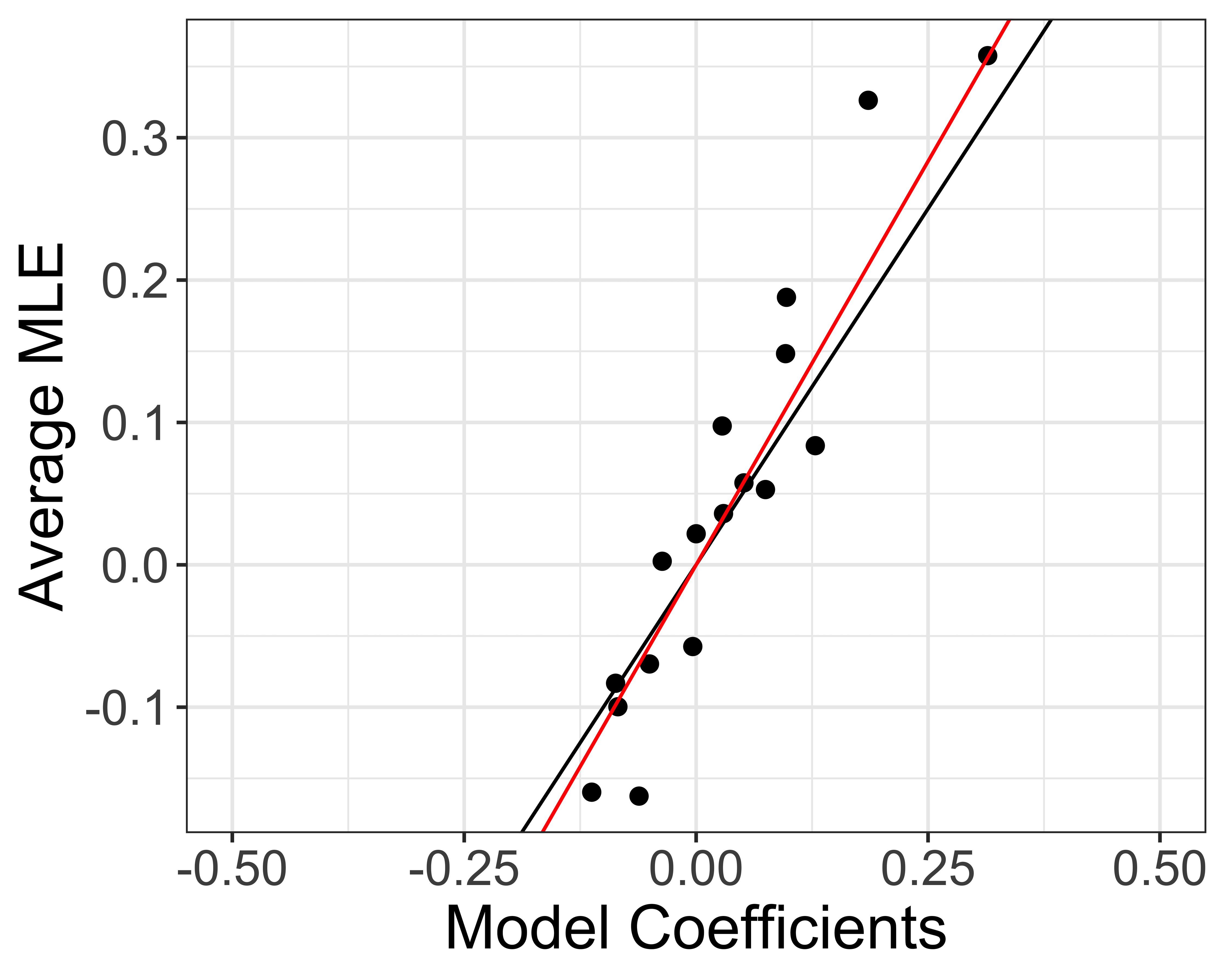}
    \hspace{1 cm}
    \includegraphics[width = 0.4\textwidth]{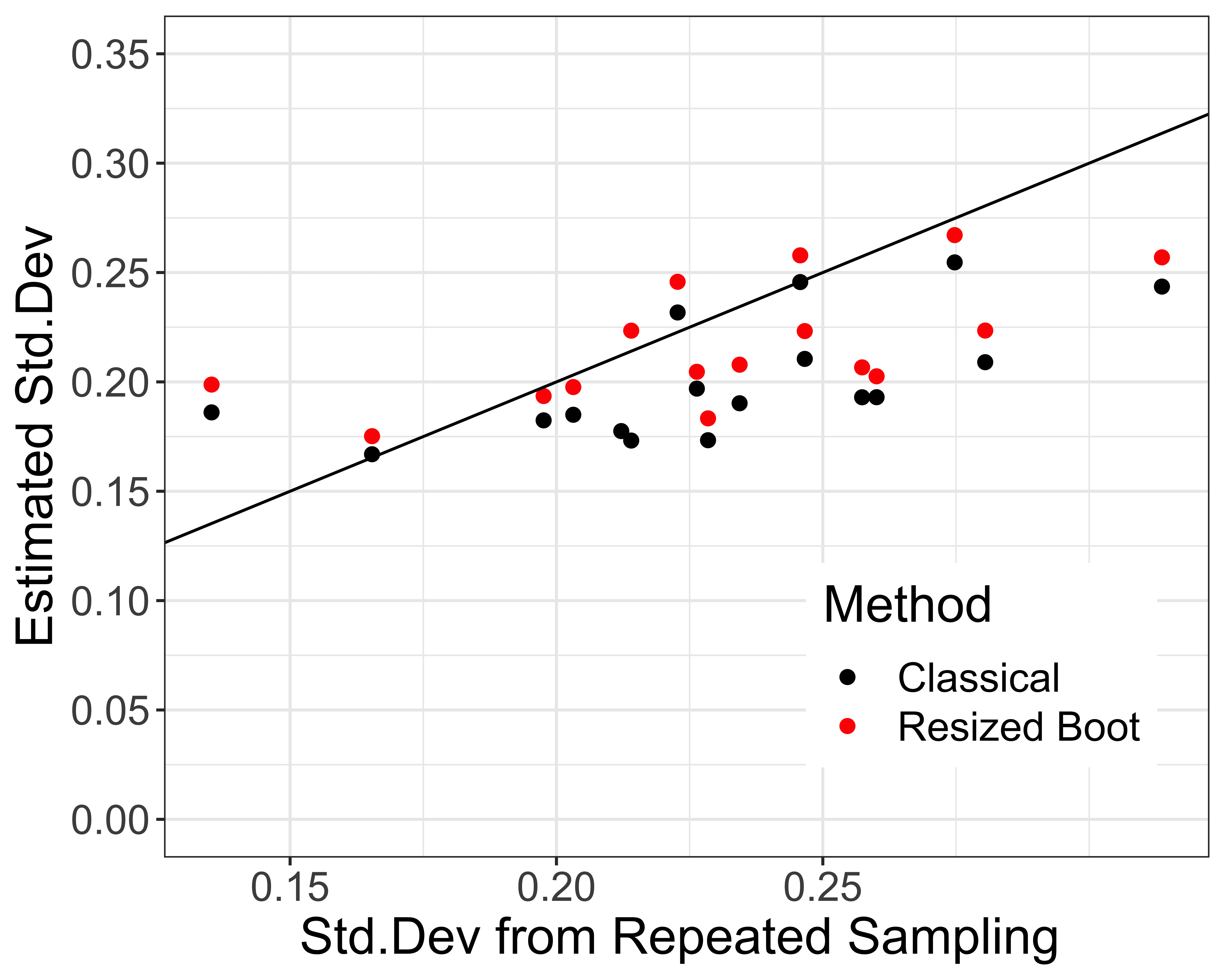}
    \caption{Bias and std.~dev.\ of the MLE. (Left) Average MLE for the
     variables versus true coefficients. The black points
      show the average MLE averaged over $B = 25$ batches. The red
      line shows the resized bootstrap estimate of the bias factor
      ($\hat{\alpha}_b = 1.13$). (Right) Average estimated standard
      deviation of the MLE for each variable versus standard deviation
      across batches. The black and red points respectively use
      classical theory and the resized bootstrap. In both plots, the black
      line is the 45 degree line.}
    \label{fig:real_bias_sd}
\end{figure}

Then, we generate $B=25$ disjoint subsamples of sample size $n = 180$ and compare classical theory and the resized bootstrap based on the estimated bias, std.~dev., and the coverage proportion of CIs. First, we examine the bias of the MLE by plotting the average of the logistic MLE estimated using each subsample versus the true coefficients (Figure \ref{fig:real_bias_sd}, Left). While the average MLEs are scattered across, their absolute magnitude is slightly larger than the true coefficients. The resized bootstrap yields an estimate $\hat{\alpha}_b = 1.13$ (red). Though this is a small adjustment, it allows the resized bootstrap to produce more accurate CI as observed in Figure \ref{fig:real_coverage}. 

Next, we plot the average estimated std.~dev.\ versus the empirical std.~dev.\ in Figure \ref{fig:real_bias_sd} (Right) calculated across batches. The resized bootstrap and the classical estimates are similar, and both methods tend to underestimate the true standard deviation. In Table \ref{tab:real_coverage}, we evaluate the proportion of variables covered in each batch as well as the coverage probability of the variable ``systolic blood pressure''. Since both methods under-estimates the std.~dev., we expect that the bootstrap provides some improvement in coverage, but does not yield correct coverage either, and this is indeed what we observe in the table. In this example, we use the large sample coefficient as a proxy for the true model coefficients, and our results suggest that when the sample size is small, while the resized bootstrap may not yield accurate coverage, it may perform better than the classical theory. 

\begin{table}[h!]
     \centering
    \begin{tabular}{|c | c c | c c|}
\hline
 Nominal &   \multicolumn{2}{c | }{I. Single variable} &  \multicolumn{2}{c|}{II. Single experiment} \\
Coverage & Classical & Resized Bootstrap & Classical & Resized Bootstrap \\
\hline
\multirow{3}{*}{}  
 95 & 87.5  (6.9)  & 91.3 (6.0) & 92.2 (1.3) & 95.0 (1.2)\\
 90  & 87.5 (6.9) & 87.0 (7.2) & 85.8 (1.6) & 88.2 (1.4)\\
 80& 83.3 (7.8) & 82.6 (8.1) & 72.3 (1.9) & 74.7 (1.7)\\
      \hline
    \end{tabular}
    \caption{Coverage probability of confidence intervals (the coverage standard deviation is between parentheses). The first
    columns report the coverage proportion for the variable ``systolic blood pressure''. The next two columns compute the proportion of variables covered in each batch and report the average over 25 batches.}
    \label{tab:real_coverage}
\end{table}

\section{Discussion}
In this paper, we demonstrated that the distribution of the MLE in
large logistic regression models depends on the distribution of the
covariates and that bootrstrap methods fail to approximate this
distribution. This is in line with previous findings concerned with
linear regression \cite{elkaroui2018,elkaroui2018boot}. To fix this
problem, we introduced a resized bootstrap, which correctly adjusts
inference.  The key is to resample from a parametric distribution
obtained by shrinking the MLE towards zero in a data-dependent
fashion, where the amount of shrinkage is informed by insights from
HDT. Resized bootstrap CIs yield correct coverage proportions for
different types of covariate distributions and types of GLMs.  Our
findings echo previous results in \cite{elkaroui2018boot,
  lopes2019spectral}; combining HDT with bootstrap
resampling methods can provide improved estimates.

We conclude with several future research questions. First, while the
resized bootstrap procedure provides a high-quality approximation to
the MLE distribution, it slightly underestimates the standard
deviation. Therefore, future research on the theoretical accuracy of
the procedure might lead to improvements in the design of the resized
MLE, for example, by adjusting the coefficients to not only match the
standard deviation of the linear predictor, but also a few higher
moments. Second, one drawback of the resized bootstrap is its
relatively high computational cost: we need to compute the MLE many
times to estimate $\gamma$ and the MLE distribution. Although a few
hundred bootstrap samples suffice to yield accurate CIs when the MLE
is approximately Gaussian, being able to reduce the computational cost
would make it even more suitable for larger datasets. Third, as
mentioned in Section~\ref{section:boot_assumption}, the resized
bootstrap is expected to accurately estimate the distribution of the
MLE for coefficients with moderate magnitudes. While the resized
bootstrap is reasonably accurate for relatively large $\beta_j$
(Appendix \ref{appendix:resizedboot_assumption}), novel insights
might further enhance it.

\subsection*{Acknowledgements}

E.~C.~was supported by the Office of Naval Research grant
N00014-20-1-2157, the National Science Foundation grant DMS-2032014,
the Simons Foundation under award 814641, and the ARO grant
2003514594.  

\printbibliography


\appendix
\section[Simulation Results in Section 5.4.2]{Additional Simulation Results}\label{appendix:additional_simulation}

This section provides additional simulation results to supplement the
findings from Section \ref{section:simulation_result}. 

Appendix
\ref{appendix:null} reports the coverage proportion of null variables
when covariates are from a multivariate $t$-distribution. Appendix
\ref{appendix:glm} shows the coverage proportion when covariates are
from a modified ARCH model and the responses from a logistic and a
probit model. Appendix
\ref{appendix:qq} shows the normal quantile plot of the MLE when the
sample size is large and covariates are i.~i.~d.\  from a Pareto
distribution.

\subsection{Coverage of a null variable}\label{appendix:null}

Table \ref{tab:single_null_mvt} reports the coverage proportion of a
null variable when the covariates follow a multivariate
$t$-distribution (see Section \ref{section:simulation_design} for the
simulation design). Coverage using either classical calculations or
the standard bootstrap is better than for a non null, compare with
Table \ref{tab:empirical_cov_single}. This is because we observed that the MLE is
unbiased when $\beta_j = 0$.

\begin{table}[!h]
    \centering
    \begin{tabular}{|c|c c | c c | c c  c c|}
        \hline
    & \multicolumn{2}{|c | }{Theoretical CI}  &\multicolumn{2}{c | }{Standard Bootstrap} & \multicolumn{4}{c|}{Resized Bootstrap}\\
      Nominal   &  & & & & \multicolumn{2}{c}{Known $\gamma$}  & \multicolumn{2}{c|}{Estimated $\gamma$} \\
      coverage & Classical & High-Dim & Parametric & Pairs & Boot-$g$  & Boot-$t$ & Boot-$g$  & Boot-$t$ \\
      \hline
\multirow{ 2}{*}{95}  & 93.1 & 93.6 & 93.4 & 95.1 & 95.5 & 95.0 & 95.6 & 95.1  \\
& (0.3) & (0.2) & (0.9) & (0.7) & (0.6) & (0.7) & (0.7) & (0.7)\\

 \multirow{ 2}{*}{90} & 87.4 & 88.1 & 88.7 & 91.2 & 90.0 & 90.2 & 89.7 & 90.1  \\
 & (0.3) & (0.3) & (1.1) & (0.9) & (0.9) & (0.9) & (1.0) & (1.0) \\
\multirow{ 2}{*}{80}  & 76.7 & 77.9 & 78.0 & 82.7 & 78.7 & 79.4 & 80.4 & 80.4 \\
&  (0.4) & (0.4) & (1.4) & (1.1) & (1.2) & (1.2) & (1.4) & (1.4) \\
\hline
    \end{tabular}
      \caption{Coverage proportion of a single \emph{null} variable ($q_j$ in Eqn.~\eqref{eq:qj}) with standard deviation between parentheses. This example uses multivariate-$t$ covariates.}
    \label{tab:single_null_mvt}
\end{table}
\subsection{Normal Quantile Plot in Section \ref{sec:small_known_parameter}}\label{appendix:qq}

\begin{figure}[h!]
    \centering
    \includegraphics[width = 0.5\textwidth]{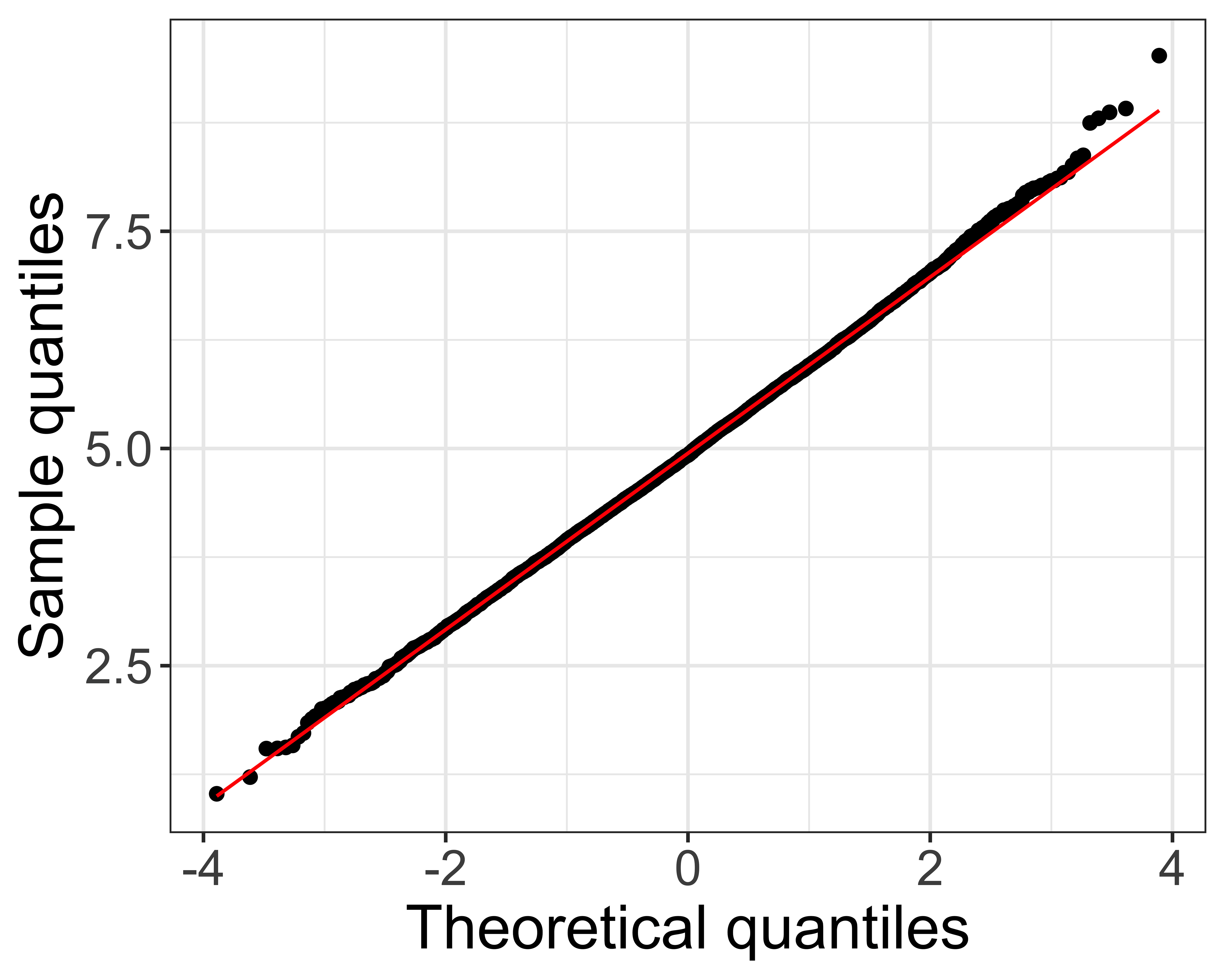}
    \caption{Normal quantile plot of an non-null MLE coordinate. The covariates are i.~i.~d.\ from a Pareto distribution. Here, $n = 4000$ and $p = 400$.
    }
    \label{fig:normal_qq_iid_large}
\end{figure}

We are in the setting of Section ~\ref{sec:small_known_parameter}, except
that $n=4,000$ and $p=400$. We see from Figure~\ref{fig:normal_qq_iid_large} that the MLE is very well approximated
by a Gaussian distribution. In contrast to the case of small samples
where the MLE has a heavy left tail (Figure~\ref{fig:boot_qqplot}), we
can say that the central limit theorem has kicked in. 

\subsection{Other Covariates and GLM}\label{appendix:glm}

We now apply the resized bootstrap to construct CIs for model
coefficients in logistic and probit regression models.  The
covariates follow our modified ARCH model. In all cases, we set
$n=4000$, and $p=400$. In probit regressions, we sample
model coefficients by first picking 50 non-null variables, and sample
the corressponding coefficients to be from an equal mixture of
$\N(3,1)$ and $\N(-3, 1)$. High-dimensional theory is available for both
logistic and probit regressions (the theory for probit is however
unpublished).

 \begin{table}[!h]
     \centering
     \begin{tabular}{|c | c|c c | c c  c c|}
         \hline
    &  & \multicolumn{2}{c | }{Theoretical CI}  & \multicolumn{4}{c|}{Resized Bootstrap}\\
   &    Nominal   &  & & \multicolumn{2}{c}{Known $\gamma$}  & \multicolumn{2}{c|}{Estimated $\gamma$} \\
   &    coverage & Classical & High-Dim & Boot-$g$  & Boot-$t$ & Boot-$g$  & Boot-$t$ \\
       \hline
\multirow{ 6}{*}{Single Null}  &  \multirow{ 2}{*}{95}  & 93.4 & 93.8 & 94.7 & 94.9 & 95.0 & 95.3 \\
&   & (0.25) & (0.24) &(0.81) & (0.70) &(0.75)  & (0.73)\\

&  \multirow{ 2}{*}{90}  & 87.6 & 88.3 & 87.8 & 88.2 & 89.2 & 88.8 \\
&    & (0.33) & (0.32) & (1.0) & (1.0) & (1.1) & (1.1) \\
 
&  \multirow{ 2}{*}{80}  & 76.6 & 77.2 & 77.5 & 77.5 & 78.0 & 78.3 \\
&  & (0.4) & (0.4) & (1.3) & (1.3) & (1.4) & (1.4) \\ 
\hline 
\multirow{ 6}{*}{Single Non-null}  &  \multirow{ 2}{*}{95}  & 81.7 & 93.0 & 93.6 & 93.7 & 93.5 & 93.5 \\
&  & (0.4) & (0.5) & (0.8) & (0.8) & (0.9) & (0.9) \\ 

&  \multirow{ 2}{*}{90}  &  72.5 & 86.8 & 87.9 & 88.4 & 87.3 & 88.5 \\
& & (0.5) & (0.3) & (1.0) & (1.0) & (1.1) & (1.1) \\ 
 
& \multirow{ 2}{*}{80}  &  58.6 & 76.2 & 77.1 & 77.2 & 77.1 & 77.1 \\
& & (0.5) & (0.4) & (1.3) & (1.3) & (1.4) & (1.5) \\ 
\hline
\multirow{ 6}{*}{Single Experiment}  &  \multirow{ 2}{*}{95}  &  92.1 & 93.7 & 94.5 & 94.8 & 94.7 & 94.9  \\

&  & (0.02) & (0.01) & (0.04) & (0.04) & (0.04) & (0.04) \\ 

 &  \multirow{ 2}{*}{90}  & 86.0 & 88.1 & 89.4 & 89.7 & 89.6 & 89.9 \\

& & (0.02) & (0.02) & (0.06) & (0.06) & (0.05) & (0.06) \\ 
 
&  \multirow{ 2}{*}{80}  &  75.0 & 77.6 & 79.5 & 79.7 & 79.7 & 79.8\\

&  & (0.03) & (0.03) & (0.08) & (0.08) & (0.08) & (0.08) \\ 
 \hline
     \end{tabular}
       \caption{ Coverage proportion of a single \emph{null} variable, single \emph{non-null} variable, and in a \emph{single-shot} experiment with standard deviation between the parentheses. This example uses modified ARCH covariates and a logistic model. Here, we use both Gaussian approximation (Column Boog-$g$) and distribution of the bootstrap MLE (Column Boot-$t$) to construct the CI. }
     \label{tab:boot_single_arch}
 \end{table}

 \begin{table}[!h]
     \centering
         \begin{tabular}{| c | c|c c | c c  c c|}
         \hline
  &   & \multicolumn{2}{c | }{Theoretical CI}  & \multicolumn{4}{c|}{Resized bootstrap}\\
  &     Nominal   &  & & \multicolumn{2}{c}{Known $\gamma$}  & \multicolumn{2}{c|}{Estimated $\gamma$} \\
   &    coverage & Classical & High-Dim & Boot-$g$  & Boot-$t$ & Boot-$g$  & Boot-$t$ \\
       \hline
\multirow{ 6}{*}{Single Null}   &  \multirow{ 2}{*}{95}  & 93.4 & 94.1 & 94.0 & 93.6 & 93.9 & 93.7  \\
 
& & (0.3) & (0.2) & (0.7) & (0.7) & (0.7) & (0.8) \\ 
&   \multirow{ 2}{*}{90}    &  87.5 & 88.5 & 89.3 & 89.2 & 88.2 & 88.7\\
  
& & (0.3) & (0.3) & (1.0) & (1.0) & (1.0) & (1.0) \\ 
 
&  \multirow{ 2}{*}{80}  & 76.9 & 78.0 & 79.2 & 79.2 & 79.1 & 79.4 \\
& & (0.4) & (0.4) & (1.2) & (1.3) & (1.3) & (1.2) \\ 
 \hline
 \multirow{ 6}{*}{Single Non-null}  & \multirow{ 2}{*}{95}  & 87.8 & 93.4 & 95.0 & 94.9 & 94.3 & 94.9  \\
& & (0.3) & (0.3) & (0.7) & (0.7) & (0.7) & (0.7) \\   

&  \multirow{ 2}{*}{90}  & 80.5 & 87.6 & 90.5 & 90.7 & 90.7 & 90.7  \\
& & (0.4) & (0.3) & (0.9) & (0.9) & (0.9) & (0.9) \\ 
 
& \multirow{ 2}{*}{80}  & 68.2 & 77.0 & 80.8 & 80.8 & 80.4 & 80.5 \\
& & (0.5) & (0.4) & (1.2) & (1.2) & (1.2) & (1.2) \\ 
\hline 
 \multirow{ 6}{*}{Single Experiment}  & \multirow{ 2}{*}{95}  &  92.0 & 93.6 & 94.6 & 94.9 & 94.8 & 95.0 \\

& & (0.02) & (0.02) & (0.04) & (0.04) & (0.04) & (0.04) \\ 

 &  \multirow{ 2}{*}{90}  & 85.6 & 88.0 & 89.6 & 89.8 & 89.7 & 89.9 \\

& & (0.02) & (0.02) & (0.06) & (0.06) & (0.05) & (0.05) \\ 
 
&  \multirow{ 2}{*}{80}  &  74.9 & 77.5 & 79.5 & 79.7 & 79.7 & 79.8\\

& & (0.03) & (0.03) & (0.08) & (0.08) & (0.07) & (0.07)  \\ 

 \hline

     \end{tabular}
      \caption{Coverage proportion of a single \emph{null} variable, single \emph{non-null} variable, and in a \emph{single-shot} experiment with standard deviation between the parentheses. This example uses modified ARCH covariates and a probit model. }
     \label{tab:boot_single_null_probit}
 \end{table}

\section{The SLOE estimator}\label{app:sloe}
The Signal Strentgh Leave-One-Out Estimator (SLOE) provides an analytic expression for estimating $\eta^2 = \lim_{n\to\infty}\var(\xnew^\top  \hat{\beta})$ where $\hat{\beta}$ is the MLE and $\xnew$ is a new observation \cite{yadlowsky2021sloe}. SLOE was developed to compute the asymptotic distribution of the logistic MLE, which depends on $\kappa = p/n$ and $\gamma^2 = \var(\xnew^\top \beta)$ and can be reparametrized to depend on $\kappa$ and $\eta$.  \cite[Proposition~2]{yadlowsky2021sloe} proves that the SLOE estimator consistently estimates $\eta$ in the high-dimensional setting. 

While SLOE was introduced for logistic regression, we generalize the
formula to other GLMs; we however do not prove consistency in this
broader setting. Define $w_i = {x}_i^\top \bvec{H}^{-1} {x}_i$, and $t_i = {x}_i^\top \hat{\beta}$,  $i=1,\ldots, n$, where $\bvec{H}$ is the
Hessian of the negative log-likelihood evaluated at the MLE
$\hat{\beta}$. Let
\[
    S_i = {X}_i^\top \hat{\beta} + q_i f'_{y_i}(t_i),
\]
where
\[
q_i = \frac{w_i}{1-w_i f''_{y_i}(t_i)}.
\]
 Above, $f_{y}(t)$ is the negative log-likelihood function when the
linear predictor is $t$ and the response is $y$,  In the case of logistic regression,
$f_y(t) = \log(1+e^{-yt}) $ for $y\in\{\pm 1\}$.

Then, we define the extended SLOE estimator to be
\begin{equation}
    \hat{\eta}^2_{\mathrm{SLOE}} = \frac{1}{n}\sum_{i} S_i^2 - \left(\frac{1}{n}\sum_{i} S_i\right)^2.
\end{equation}
Here, $S_i$ approximates $x_i^\top \hat{\beta}_{(i)}$, where $\hat{\beta}_{(i)}$ is the MLE computed without using the $i$th observation. Since $x_i$ is independent of $\hat{\beta}_{(i)}$, the variance of $S_i$ approximates $\var(\xnew^\top \hat{\beta})$.

\section{Simulated example when the coefficients are sparse}\label{appendix:resizedboot_assumption}
To study the accuracy of our method for large coefficients, we use a
simulated example where there are only 10 non-null variables whose
coefficients have equal magnitudes, which equals to 10, and $\pm 1$ signs with equal
probability. Here, $\tau_j|\beta_j|/\gamma \approx 0.32$ (in this case, $\tau^2_j=\var(X_j|X_{-j}) = 1/p = 0.025$).

We first examine the bias and variance of the MLE
(Table~\ref{tab:sparse_bias_variance}). We report the average bias of
all of the non-null variables in $N=10,000$ repeated experiments
(Column Empirical) and the estimated bias using high-dimensional
theory (Column High-Dim Theory) and the bootstrap (Column Resized
Bootstrap). We observe that the high-dimensional theory slightly
under-estimates the bias, with a relative error of about 1\%, whereas
the bootstrap estimates are more accurate. We then study the
std.~dev.\ of the MLE and report the average std.~dev.\ of all of the
null variables (Table~\ref{tab:sparse_bias_variance}, Row std.~dev.\
(null)) and the non-null variables
(Table~\ref{tab:sparse_bias_variance}, Row std.~dev.\ (non-null)). As in
Table~\ref{tab:bias_variance}, the variance of the non-null variables
are higher than that of the null variables. On the other hand, unlike
Table~\ref{tab:bias_variance}, the high-dimensional theory
underestimates the variance of both the null and nonnull variables
(recall that the covariates are not Gaussian). 
Lastly, the resized bootstrap method using either a known
or estimated $\gamma$, also slightly underestimates the variance of
the non-null variables. It is however more accurate than HDT with a
relative error below 1\%. This shows that the bootstrap is reasonably
accurate for large coefficients. 

\begin{table}[h!]
  
    \centering
    \begin{tabular}{|l| c c c c c |}
    \hline
& & Classical & High-Dim &  \multicolumn{2}{c |}{Resized Bootstrap} \\
&      Empirical & Theory  & Theory & Known $\gamma$ & Estimated $\gamma$ \\
      \hline
Bias   & 1.15 & - &  1.14 & 1.15  &  1.15 \\
Std.~dev.\ (null)& 0.98 & 0.92 & 0.93 & 0.98 & 0.98 \\
Std.~dev.\ (non-null) & 1.05 & 0.97 & 0.93 & 1.02 & 1.03\\
\hline
    \end{tabular}
    \caption{ The average bias and variance of the MLE when the coefficients are large ($\tau_j\beta_j/\gamma\approx 0.32$). The second and third row report average std.~dev.\ for a single null or non-null variable. In this simulation, the covariates are from a modified ARCH and the responses are from a logistic regression. The resized bootstrap estimates are averaged in $N=1,000$ simulations.}
    \label{tab:sparse_bias_variance}
\end{table}

We next study the coverage probability of confidence intervals by
computing the average coverage proportion of the null and non-null
variables. Unsurprisingly, the HDT (Table \ref{tab:sparse_coverage},
Column High-Dim Theory) undercovers both for the null and non-null
variables, and the coverage proportions are less accurate for non-null
variables. The resized bootstrap using the correct $\gamma$ (Table
\ref{tab:sparse_coverage}, Column Known $\gamma$) slighty undercovers
but the coverage is closer to the nominal coverage. Interestingly, the
resized bootstrap with an estimated $\hat \gamma$ nearly achieves
nominal coverage for both null and non-null variables at every
considered significance level. In contrast, classical theory CI significantly undercovers the true coefficient because the classical theory does not account for the bias of the MLE.

Our findings here agree with our hypothesis that the resized bootstrap
is less accurate when the coefficient is large. At the same time, our
results are reassuring because they suggest that the resized bootstrap
is reasonably accurate for relatively large coefficients.

\begin{table}[h!]    
    \centering
    \begin{tabular}{| l|c c  c  c c|}
        \hline
&Nominal & Classical & High-Dim &  \multicolumn{2}{c |}{Resized Bootstrap}  \\
Variable & Coverage  & Theory & Theory & Known $\gamma$ & Unknown $\gamma$\\
\hline
\multirow{6}{*}{Null} 
& \multirow{2}{*}{95} & 93.4 & 93.5 & 95.2 & 95.6\\
& & (0.2) & (0.3) & (0.6) & (0.6)\\
  & \multirow{2}{*}{90} & 87.9 & 88.0 & 88.9 & 89.6 \\
& & (0.3)& (0.3) & (0.9) & (0.9)\\
& \multirow{2}{*}{80} &  77.1 & 77.3 & 78.8 & 78.8 \\
& &(0.4)& (0.4) & (1.2) & (1.2)\\
\hline
\multirow{6}{*}{Non-null} & 
\multirow{2}{*}{95} &  64 & 91.4 & 94.4 & 94.7 \\
& & (0.5) & (0.3) & (0.7) & (0.6)\\
  & \multirow{2}{*}{90} & 52 & 84.5 & 88.8 & 89.0  \\
& & (0.5) & (0.4) & (0.9) & (0.9)\\
& \multirow{2}{*}{80} & 38.0 & 73.8 & 79.4 & 78.8 \\
& & (0.5) & (0.4) & (1.2) & (1.2)\\
\hline
    \end{tabular}
   \caption{Average coverage proportion of a single null or nonnull variable. The std.~dev.\  is reported inside the parentheses. We use bootstrap-$t$ confidence intervals in this example.}

    \label{tab:sparse_coverage}
\end{table}

\end{document}